\documentclass[12pt,letterpaper]{article}

\usepackage[margin=1in]{geometry}

	\usepackage{amsmath}
	\usepackage{amsthm}
	\usepackage{amssymb}
	\usepackage{amsfonts}
	\usepackage{mathtools}
	\usepackage{booktabs}
	\usepackage{array}
	\usepackage{calc}
        \usepackage{multirow} 
	\usepackage{graphicx}
	\usepackage{float}
	\usepackage{rotating}
	\usepackage{tikz}
	\usepackage{caption}
	\usepackage[labelformat=simple]{subcaption}
	
	\graphicspath{ {input/} }
	
	\usepackage{tabularx}
	\usepackage{longtable}
	\usepackage{multirow}
	\usepackage{makecell}
	\usepackage{threeparttable}

	\usepackage{xcolor}
	\usepackage{colortbl}
	\definecolor{greenlight}{RGB}{212, 237, 218}
	\definecolor{greenmedium}{RGB}{195, 230, 203}
	\definecolor{greendark}{RGB}{179, 217, 192}
	\definecolor{redlight}{RGB}{248, 215, 218}
	\definecolor{redmedium}{RGB}{245, 198, 203}
	\definecolor{reddark}{RGB}{241, 176, 183}
	\definecolor{baseline}{RGB}{233, 236, 239}

	\usepackage[authoryear,round]{natbib}
	
	\usepackage[english]{babel}
	\usepackage[T1]{fontenc}
	\usepackage[utf8]{inputenc}
	\usepackage{setspace}
	\usepackage{soul}
	\usepackage{comment}
	\PassOptionsToPackage{hyphens}{url}
	\usepackage{hyperref}
	\hypersetup{
		urlcolor=blue,
		citecolor=blue,
		linkcolor=blue,
		colorlinks=true,
	}

	\newcommand{\GG}[1]{}

	\newcolumntype{L}[1]{>{\raggedright\let\newline\\\arraybackslash\hspace{0pt}}m{#1}}
	\newcolumntype{C}[1]{>{\centering\let\newline\\\arraybackslash\hspace{0pt}}m{#1}}
	\newcolumntype{R}[1]{>{\raggedleft\let\newline\\\arraybackslash\hspace{0pt}}m{#1}}

\title{Macroeconomic Forecasting and Machine Learning}

\author{Ta-Chung Chi\footnote{Carnegie Mellon University. Email: \href{mailto:tachungc@alumni.cmu.edu}{tachungc@alumni.cmu.edu}} \and Ting-Han Fan\footnote{Princeton University. Email: \href{mailto:tinghanf@alumni.princeton.edu}{tinghanf@alumni.princeton.edu}} \and Raffaele M. Ghigliazza\footnote{Princeton University. Email: \href{mailto:raffaele.ghigliazza@gmail.com}{rghiglia@alumni.princeton.edu}} \and Domenico Giannone\footnote{Johns Hopkins University and CEPR. Email: \href{mailto:domenico.giannone@jhu.edu}{domenico.giannone@jhu.edu}} \and Zixuan (Kevin) Wang\footnote{Harvard University. Email: \href{mailto:zixuanwang@alumni.harvard.edu}{zixuanwang@alumni.harvard.edu}.
\newline We thank Christian Haefke, Davide Pettenuzzo, Frank Diebold, Mike McCracken, Raffaella Giacomini, and Todd Clark for comments and suggestions.}}

\date{}

\begin{document}
\maketitle
\thispagestyle{empty}
\begin{abstract}
We forecast the full conditional distribution of macroeconomic outcomes by systematically integrating three key principles: using high-dimensional data with appropriate regularization, adopting rigorous out-of-sample validation procedures, and incorporating nonlinearities. By exploiting the rich information embedded in a large set of macroeconomic and financial predictors, we produce accurate predictions of the entire profile of macroeconomic risk in real time. Our findings show that regularization via shrinkage is essential to control model complexity, while introducing nonlinearities yields limited improvements in predictive accuracy. Out-of-sample validation plays a critical role in selecting model architecture and preventing overfitting.
\end{abstract}

\noindent \textbf{JEL Codes}: C22, C52, C53, C55. \\
\noindent \textbf{Keywords}: Deep Neural Network, Quantile Regression, Unemployment at Risk, Regularization, Big Data, Out-of-Sample Validation.
\vskip3cm
\newpage
\onehalfspacing
\setcounter{page}{1}

\section{Introduction}

Forecasting has undergone a profound transformation in the 21st century, driven by advancements in methodology, computational power, and data availability. The origins of this transformation can be traced back to the late 1990s, when the field of economics began grappling with the challenges and opportunities presented by what is now termed ``Big Data''—the availability of large datasets with numerous predictors. This period marked the emergence of systematic efforts to develop tools capable of addressing the high-dimensional nature of these datasets with seminal contributions by Frank Diebold, Mario Forni, Marc Hallin, Marco Lippi, Lucrezia Reichlin, Jim Stock, and Mark Watson \citep[see][]{Reichlin1998, Watson1998, Diebold1998}. They laid the foundation for a new era of forecasting, as presented at the World Meeting of the Econometric Society in the summer of 2000 \citep[for a survey see][]{DemolEtal2017,Diebold2021}.

Since these early contributions, the field of macroeconomic forecasting has experienced significant progress. Many of these advances were developed within macroeconometrics, while others originated independently in different disciplines, often leading to valuable cross-fertilization. Taken together, these developments fall under the broader umbrella of machine learning (ML), which has become one of the most influential ideas of the 21st century, reshaping empirical modeling across science, industry, and society. Motivated by the success of these methods in other fields, economists have in some cases rediscovered and re-emphasized key ideas, giving them renewed momentum and sparking broader interest across the profession. At times, however, the connections between modern machine learning and the macroeconomic forecasting tradition are overlooked. This can create confusion about what is genuinely new: some economists, familiar with the history of macroeconomic forecasting, may see ML as ``reinventing the wheel,'' while others may regard it as an entirely new paradigm. This disconnect can hinder communication, mutual understanding, and trust, even within the same field, by obscuring both novelty and continuity in today's forecast innovations.

At a high level, three key ideas now underpin modern forecasting approaches: using high-dimensional data with appropriate regularization, adopting rigorous out-of-sample procedures for both testing and validation, and incorporating nonlinearities. Most of these elements have deep roots in the theory and practice of macroeconomic forecasting. In addition, over the past decade, there has been growing attention to predicting the full distribution of risks, reflecting increased interest in tail events and macroeconomic uncertainty. 

In this paper, we take stock of these developments and synthesize them into a unified framework for predicting macroeconomic risk. In summary, our results show that the model can effectively capture emerging macroeconomic vulnerabilities—especially those related to downside risks—by extracting information from high-dimensional data through a flexible and general predictive framework. We find little variation in upside risk but substantial time variation in downside risk. Thanks to validation-based selection and appropriate regularization, the high dimensionality of the data and the complexity of deep neural networks do not lead to overfitting issues. While shrinkage proves essential in this context, allowing for nonlinearities provides only limited gains in predictive accuracy.

The rest of the paper is organized as follows. Section \ref{sec:lit} provides a summary of the evolution of the literature on macroeconomic forecasting. Section \ref{sec:math} introduces the problem formally. In Section \ref{sec:empirical}, we apply the framework to forecasting the unemployment rate one month ahead. Section \ref{sec:results} discusses the results, and Section \ref{sec:conclusion} concludes.

\section{Literature}
\label{sec:lit}
In this paper, we use the term ``Big Data'' as it is understood in macroeconomics: a setting in which model complexity is high relative to the number of observations \citep{DemolEtal2017, GiannoneEtal2021}. In such environments, even simple linear models can overfit, especially when the number of predictors approaches or exceeds the number of observations. In-sample performance often fails to translate into reliable out-of-sample forecasting accuracy, as the high parameter-to-observation ratio leads to poor generalization. Two major strategies have emerged to address these challenges: dimensionality reduction via principal components and regularization through shrinkage. \citet{SW2002, SW2005JBES}, \citet{ForniEtal2000, ForniEtal2005}, and \citet{West2003} introduced dynamic factor models for large datasets and showed that these models provide an effective way to summarize the information in many predictors. For a comparative overview of dynamic factor models in macroeconomic forecasting, see \citet{SW2006}, \citet{DagostinoGiannone2012}, \citet{SW2016}, and \citet{DozFuleky2020}. 

In parallel, shrinkage methods such as ridge regression, Lasso, and spike-and-slab regressions were developed to penalize complexity and mitigate overfitting. \citet{DemolEtal2008} introduced shrinkage methods in the context of forecasting with many predictors. They showed that ridge regression (which uses an $\ell_2$ penalty) is asymptotically equivalent to dynamic factor models when predictors are strongly correlated, a common feature of macroeconomic datasets. This literature and the connection between penalty functions and informative priors have led to the widespread adoption of Bayesian inference in large-dimensional systems, as Bayesian inference provides both computational tractability and principled uncertainty quantification, particularly through the development of Large Bayesian VARs \citep{BanburaEtal2010,Koop2013JAE,GiannoneEtal2015}.

\citet{DemolEtal2008} also found that Lasso (which uses an $\ell_1$ penalty) achieves similar predictive performance, but the selection of predictors is unstable. \citet{BianchiEtal2022} studied elastic net (combining of $\ell_1$ and $\ell_2$ penalties) while \citet{GiannoneEtal2021} considered spike-and-slab priors (combining $\ell_0$ and $\ell_2$ penalties). Sparsity-inducing methods have gained significant traction in statistics and have been successfully applied across a variety of disciplines, from seismography and medical imaging to astronomy and signal processing \citep[e.g.,][]{SantosaSymes1986, DaubechiesDefriseDeMol2004, Tibshirani1996, BickelEtal2009, HastieTibshiraniWainwright2015}. In economics, these methods have proven effective in financial applications—for instance, in constructing sparse and stable portfolios \citep{BrodieEtAl2009}—as well as in micro-econometric settings, particularly for high-dimensional treatment effect estimation and instrument selection \citep{BelloniEtal2013}. However, in macroeconomic forecasting, the use of sparsity-based techniques has remained relatively limited. Researchers have instead relied more heavily on ridge regression and factor model approaches. This reflects the fact that macroeconomic datasets typically exhibit strong co-movement among predictors, which makes sparsity less relevant and favors dense representations that capture common underlying dynamics \citep[see][]{DemolEtal2017, GiannoneEtal2021}.

During the 2000s, most forecasting applications remained grounded in linear models, given their empirical success and interpretability. Nevertheless, researchers explored nonlinear methods. Notable early contributions include \citet{White1988,white1989learning,DieboldNason1990,SwansonWhite1997,HaefkeEtal1997,White2006}, who employed neural networks for macroeconomic forecasting. Yet empirical evidence during that period suggested that nonlinearities yielded limited improvements over linear benchmarks  \citep[see also][]{HevaviEtal2004}. \citet{White1988} and \citet{DieboldNason1990} found similar results for predictions of stock prices and exchange rates, respectively. Combined with the computational burden of estimating nonlinear models in high dimensions, this led to a continued emphasis on linear approaches. Recent increases in computational power and algorithmic efficiency have revitalized interest in nonlinear methods, especially deep learning. Models such as regression trees and deep neural networks (DNNs) can now be estimated and applied at scale, enabling richer approximations of complex functional relationships \citep{LenzaEtal2023,ClarkEtal2023,CarrieroEtal2024,HauzenbergerEtal2025}. 

Another major development was the growing role of out-of-sample (OOS) validation. While OOS evaluation has long been a cornerstone of econometric forecast comparison and hypothesis testing, it became especially prominent in high-dimensional settings. Seminal contributions by \citet{DieboldMariano1995}, \citet{West1996}, \citet{White2000}, \citet{ClarkMcCraken2001},\citet{GiacominiKomunjer2005}, \citet{GiacominiWhite2006}, \citet{AmisanoGiacomini2007}, \citet{ClarkWest2007}, \citet{GiacominiKomunjer2005}, and \citet{RossiSekhposyan2019} established OOS validation as a key tool for evaluating forecast accuracy \citep[see also the survey in][]{ClarkMcCracken2013,GiacominiRossi2013,Komunjer2013}. The shift toward OOS reasoning occurred in tandem with the Big Data revolution, where model complexity made in-sample fit unreliable. A key early contribution was \citet{SW1996, SW1999}, who pioneered the systematic use of OOS model comparison in high-dimensional macroeconomic forecasting. In this context, in-sample criteria often fail to capture genuine predictive performance. Machine learning (ML) has pushed this logic further by embedding out-of-sample evaluation into every stage of model development. In modern ML workflows, performance on held-out data is not just used for testing, but also for selecting hyperparameters and designing model architectures. The logic of out-of-sample evaluation for hyperparameter tuning and model selection is deeply rooted in classical tools such as cross-validation and hierarchical Bayesian inference. These approaches offer a principled framework for regularization that naturally incorporates out-of-sample validation into the estimation process. For a discussion of the connection between hierarchical Bayes and out-of-sample validation for model selection in a Big Data setting, see \citet{GiannoneEtal2015, GiannoneEtal2021}.

While earlier work in macroeconomic forecasting focused primarily on point predictions, the past decade has seen growing attention to forecasting the full distribution of risks. \citet{AdrianEtal2019} show that macroeconomic risk—especially on the downside—varies substantially over time and can be predicted using financial and macroeconomic indicators. Their original analysis focused on U.S. GDP growth and introduced the concept of growth-at-risk to summarize the evolving distribution of future outcomes. This perspective has since been extended to other countries and macroeconomic variables, including inflation and unemployment \citep[e.g.,][]{Figueres_Jarocinski_2020, AABG2021, kiley, Amburgey2023, boyarchenko2023outlook, Chernis2023JAE,Loria2024Inflation}. Although this literature often incorporates Big Data, typically by using indexes or factors extracted from large panels of macroeconomic and financial variables, it has not fully exploited recent advances in machine learning, particularly the flexible incorporation of high-dimensional predictors, nonlinearities, and rigorous out-of-sample validation. Several recent studies represent important steps toward a more systematic integration of these components. For example, \citet{LenzaEtal2023} use quantile regression forests to forecast the distribution of euro area inflation, capturing nonlinear interactions between inflation and financial conditions.  \citet{Hengge2019} integrates multiple measures of financial conditions and economic and political uncertainty to forecast economic growth. The \citet{IMF2024} applies deep neural networks to assess the predictive power of financial conditions and uncertainty for macroeconomic risk. \citet{furceri2024global} use many macroeconomic and political factors to predict both the central trajectory of public debt and the uncertainty surrounding it across more than one hundred countries. There is also a long tradition of forecasting risk in finance: \citet{ForesiPeracchiJasa2005} forecast the cumulative distribution function of excess returns conditional on a broad set of predictors and \citet{CrumpEtal2024} combine density forecasts of stock returns to improve overall forecast accuracy. Out-of-sample evaluation beyond point forecasts also has a long tradition in macroeconomics. \citet{AmisanoGiacomini2007} study the evaluation of the accuracy of density forecasts. \citet{DieboldEtal1999,RossiSekhposyan2019} develop methods to assess the calibration of density forecasts. \citet{GiacominiKomunjer2005} consider the accuracy of quantile predictions \citep[see][for a survey.]{Komunjer2013}. \cite{JoreEtal2010,ConflittiEtal2015} study how to combine density forecasts to improve out-of-sample accuracy.

Our contribution builds on these advancements by systematically combining all key elements—regularization, nonlinearity, out-of-sample validation and testing, and full predictive distribution—into a unified high-dimensional forecasting framework. We adopt a nested modeling approach that not only integrates these components, but also allows us to assess the relative importance of each in shaping predictive performance.

The forecasting exercise is deliberately stylized and does not capture the full complexity of real-time macroeconomic data, nor significant nonlinearities. While we carefully avoid look-ahead bias by using a recursive design that mimics real-time information availability during model training, validation, and testing, we do not incorporate several key real-time features. In particular, our analysis omits the non-synchronicity of data releases, mixed-frequency observations, and data revisions—dimensions emphasized in the nowcasting literature \citep{GRS2008,BanburaEtal2013,Luciani,CascaldiEtal2024}. Nonetheless, the core lessons from our analysis remain relevant, as nowcasting models often rely on similar modeling principles and estimation strategies, particularly the use of regularization to handle big data.

\section{Machine Learning}
\label{sec:math}

We are interested in estimating the following object: 
\[
P(Y_{T+h} < y \mid X_T = x_T) = F_{Y_{T+h} \mid X_T = x_T}(y)
\]
The function \( F_{Y_{T+h} \mid X_T = x_T}(y) \) is the conditional cumulative distribution function (CDF) of \( Y_{T+h}\) given the predictors \( X_T = x_T \). Here, \( X_T \) is a \( k \)-dimensional vector of predictors, with \( x_T \) being the observed realization of \( X_T \), and \( k \) is large, meaning the model must handle a high-dimensional feature space.

\subsection{Conditional Quantile Function}

Quantile prediction has a long tradition in econometrics and statistics, where it has been used extensively to model conditional distributions beyond the mean \citet[see][for a comprehensive survey.]{Komunjer2013}.

Instead of estimating the cumulative distribution function (CDF) directly,\footnote{The direct estimation of the CDF can be performed using distribution regression, as developed by \citet{ForesiPeracchiJasa2005}.}  we estimate its inverse—the quantile function—defined as
\[
Q_{Y_{T+h} \mid X_T = x_T}(\tau) = y \quad \text{where} \quad \tau = F_{Y_{T+h} \mid X_T = x_T}(y).
\]
That is, the \(\tau\)-th conditional quantile of \(Y_{T+h}\) given \(X_T = x_T\) is the value \(y\) such that the probability of observing an outcome below \(y\) is \(\tau\).

Quantiles can be equivalently characterized as the solution to an optimization problem. Specifically, the \(\tau\)-th quantile choose the optimal \(q\)  that minimizes an asymmetric linear loss function known as the \textit{quantile loss}, or \textit{pinball loss}:

\begin{equation*}
L_{\tau}(y_{T+h}, q) = \tau \cdot \max(y_{T+h} - q, 0) + (1 - \tau) \cdot \max(q - y_{T+h}, 0).
\end{equation*}

\begin{figure}[H]
    \centering
    \includegraphics[width=0.85\linewidth]{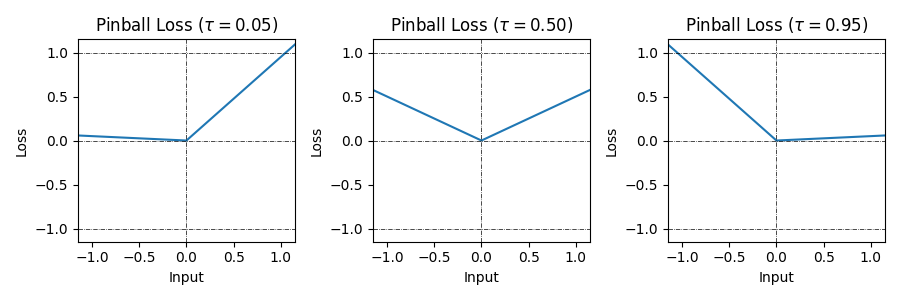}
    \caption{Pinball loss functions for quantiles \(\tau = 0.05\), \(\tau = 0.5\), and \(\tau = 0.95\)}
    \label{fig:pinball}
\end{figure}

This loss function penalizes under-predictions and over-predictions differently depending on the chosen quantile $\tau$. For \(\tau = 0.5\), the pinball loss function is symmetric: it assigns equal weight to positive and negative errors, and the resulting V-shape has equal slopes on both sides. However, when \(\tau \neq 0.5\), the function becomes \emph{tilted}, introducing asymmetry in the penalization of forecast errors. When \(\tau < 0.5\), the loss function assigns \emph{more weight to negative errors}—that is, when the predicted quantile \(q\) exceeds the actual outcome \(y_{T+h}\). The slope on the left side of the V is steeper, encouraging the model to shift the quantile downward so that a smaller proportion of observations fall below the estimated quantile line. When \(\tau > 0.5\), \emph{positive errors}—when the predicted quantile is below the actual value—are penalized more. The right-hand slope becomes steeper, shifting the estimated quantile upward so that a larger proportion of observations fall below it. These asymmetries are clearly visualized in Figure~\ref{fig:pinball}, which plots the pinball loss for \(\tau = 0.05\), \(\tau = 0.5\), and \(\tau = 0.95\), respectively. The loss is also known as \emph{lin-lin loss} since it is linear on each side of the origin, with potentially different slopes \cite[see][]{ChristoffersenDiebold1997}.  It also goes by the names \emph{check loss} or \emph{tick loss}, due to its resemblance to a check mark when asymmetric, with a steeper slope on one side.

This tilting of the loss function reflects the fundamental goal of quantile estimation: to find the value \(q\) such that approximately \(\tau\) of the conditional distribution lies below it. As Figure~\ref{fig:pinball} shows, the shape of the loss function adjusts with \(\tau\), becoming increasingly asymmetric as the quantile moves away from the median.

The case \(\tau = 0.5\) can play a special role since it corresponds to the \emph{mean absolute deviation}, a robust alternative to the squared loss. In contrast to the quadratic loss, which increases the weight of large errors quadratically, the pinball loss grows linearly with the size of the error. This makes it more robust to outliers and skewed distributions. Historically, the quadratic loss (i.e., squared error) has been widely used due to its analytical convenience in settings like OLS or ridge regression. However, with modern computational resources, estimating quantiles via the pinball loss is no longer burdensome. In our analysis, we use the median prediction for point forecasts and report results under both the \(\ell_1\) and \(\ell_2\) losses for robustness.

The quantile function is estimated by minimizing this loss. We treat the quantile function \( Q_{Y_{T+h} \mid X_T = x_T}(\tau) \) for each quantile level \( \tau \) as a general function of the predictors \( x_T \). We approximate this quantile function in a parametric form, where it depends on the predictors \(x_T\), on some parameters \(\theta, \psi \), denoted as $$ Q_{Y_{T+h} \mid X_T = x_T}(\tau) \approx f_{\psi,\tau}(x_T; \theta).$$

This function will be estimated using a deep neural network. The parameters $\psi$ specify the architecture and $\theta$ the trainable parameters. As described later, to regularize the estimation, we will augment the quantile loss with a penalty on the model's parameters.

The computation of the quantiles was once challenging computationally, up to a few decades ago, and there is a large literature using linear programming to compute conditional quantiles. Nowadays, we are able to solve these problems in high dimensions and also for non-linear relationships.

\subsection{Deep Neural Networks}

This section introduces Deep Neural Networks (DNNs) and their application to nonlinear prediction problems. We begin by outlining the general structure of a DNN, which consists of an input layer, one or more hidden layers, and an output layer. The core mechanism involves stacking multiple linear transformations interleaved with nonlinear `activation' functions\footnote{Mimicking the thresholding behavior in biological neurons}. For simplicity and clarity, we then illustrate specific cases with zero, one, and two hidden layers. These cases help relate DNNs to familiar econometric models such as linear regression and reduced-rank regression. Finally, we discuss the choice of activation function, focusing on the Leaky ReLU, a widely used and flexible nonlinear transformation.

Deep Neural Networks (DNNs), which extend classical Multi-Layer Perceptrons (MLPs), have become the foundation of modern machine learning. Originally introduced to overcome the limitations of shallow architectures, DNNs rose to prominence as advances in computation and optimization made training deep architectures feasible. Although the concept of MLPs dates back to the 1950s and 1960s---originating from early models like Rosenblatt's perceptron---the breakthrough came in the 1980s with the introduction of the backpropagation algorithm. The term "deep" entered common use in the mid-2000s, when researchers showed that networks with many hidden layers could be trained effectively, notably in work by \cite{rumelhart1986learning} and \cite{hinton2006fast}.

\subsubsection{DNN Structure}
\begin{figure}[H]
    \centering
    \includegraphics[width=0.65\linewidth]{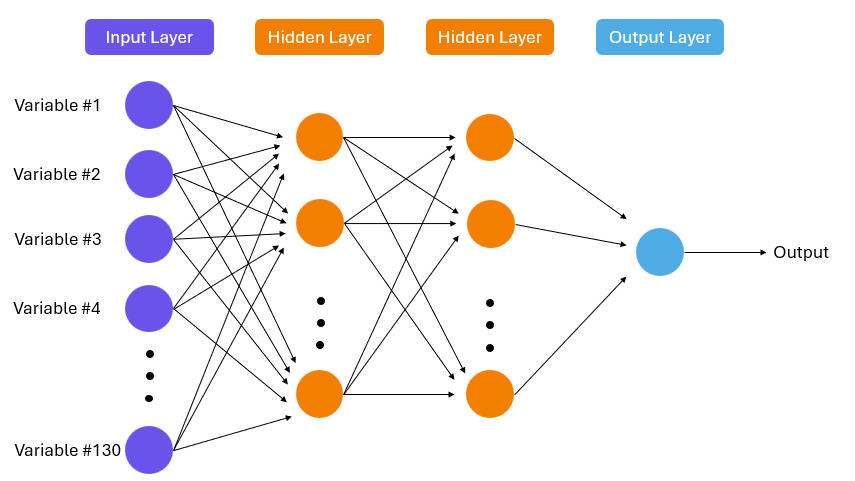}
    \caption{Schematic representation of a Deep Neural Network (DNN). An example with an input layer with 130 features, followed by two hidden layers and a scalar output.}
    \label{fig:dnn}
\end{figure}

Graphically, the structure of a DNN is illustrated in Figure~\ref{fig:dnn}. The figure shows a typical architecture consisting of an input layer, two hidden layers, and an output layer. Here, the input layer takes 130 features (which might represent macroeconomic indicators or other predictors). The hidden layers, assemble the inputs into the layer and apply nonlinear transformations, enabling the network to learn complex representations and shown to be `universal' (continuous) function approximators \cite{hornik1989multilayer}. The output layer produces a scalar prediction. We define each of these components mathematically below.

Let \( x_t \in \mathbb{R}^k \) denote the input vector at time \( t \), consisting of \( k \) features. A deep neural network maps this input into an output \( y_t \in \mathbb{R}^n \) by passing it through a sequence of transformations. The architecture consists of an input layer, a series of hidden layers, and an output layer. Each hidden layer applies a non-linear activation function to a linear transformation of its input.

The input layer simply takes the feature vector \( x_t \) and feeds it into the network:
\[
x^{(0)} = x_t \in \mathbb{R}^k.
\]
This vector represents the raw input data, for instance, a set of macroeconomic indicators or other predictors.

Each hidden layer performs two operations: a linear transformation followed by a non-linear function, the `activation' function. Let \( x^{(i)} \in \mathbb{R}^{r_i} \) denote the output of the \( i \)-th hidden layer ($i>0$). The transformation is given by:
\[
x^{(i)} = a_i(W^{(i)} x^{(i-1)} + b^{(i)}),
\]
where:
\begin{itemize}
    \item \( W^{(i)} \in \mathbb{R}^{r_i \times r_{i-1}} \) is the weight matrix,
    \item \( b^{(i)} \in \mathbb{R}^{r_i} \) is the bias vector,
    \item \( a_i(\cdot) \) is the activation function.
\end{itemize}

The final layer takes the output of the last hidden layer and maps it to a prediction. For quantile regression, we assume a linear output layer:
\[
Q_{Y_{t+h} \mid X_t = x_t}(\tau) \approx \gamma^\top x^{(d)} + c,
\]
where \( x^{(d)} \) is the scalar output of the last hidden layer, \( \gamma \) is the weight vector, and \( c \) is a constant.

\subsubsection{Examples}

While the general structure of a DNN is straightforward, the recursive notation used to describe multiple layers can appear cumbersome. To build intuition and relate the architecture to familiar forecasting models, we now present a sequence of examples. We begin with the simplest case involving only the output layer, and then consider models with one and two hidden layers.

\noindent \textbf{Example 0}: If the network has no hidden layers (\( d = 0 \)), the model reduces to:
\[
Q_{Y_{t+h} \mid X_t = x_t}(\tau) \approx \gamma^\top x_t + c,
\]
which is the familiar linear quantile regression model. If a penalty is added on \( \gamma \), the model corresponds to a ridge-regularized quantile regression.

\noindent \textbf{Example 1}: With one hidden layer of dimension \( r_1 \), the network becomes:
\[
x^{(1)} = a_1(W^{(1)} x_t + b^{(1)}),
\]
\[
Q_{Y_{t+h} \mid X_t = x_t}(\tau) \approx \gamma^\top x^{(1)} + c.
\]
If the activation function \( a_1 \) is linear, this corresponds to a reduced-rank quantile regression.

This model might resemble unsupervised dimensionality reduction like PCA. However, in the latter case, the projection coefficients \( W^{(1)} \) are chosen to maximize the explained variance of the predictors. Instead, here the projection is optimized directly to minimize prediction error. This makes the approach closer to reduced-rank regression than to unsupervised dimensionality reduction.

\noindent
\textbf{Example 2}: With two hidden layers, the transformations become:
\[
x^{(1)} = a_1(W^{(1)} x_t + b^{(1)}),
\]
\[
x^{(2)} = a_2(W^{(2)} x^{(1)} + b^{(2)}),
\]
\[
Q_{Y_{t+h} \mid X_t = x_t}(\tau) \approx \gamma^\top x^{(2)} + c.
\]
Again, if both activation functions are linear, the model reduces to a multi-layer generalization of reduced-rank regression.

\subsubsection{Leaky ReLU}
The activation function introduces nonlinearity into the model, allowing the network to approximate complex relationships. The list of activation functions is long, from piecewise linear like the ReLU (Rectifier Linear Unit) to leaky ReLU, to sigmoidal, hyperbolic tangent, and so on. We use the piecewise linear \textbf{Leaky ReLU}:

\[
a(x; \alpha) = \begin{cases}
    x, & x \geq 0 \\
    \alpha \cdot x, & x < 0
\end{cases}
\]
where the hyperparameter \( \alpha \in [0,1] \) controls the slope on the negative side.

This function spans a range of behaviors:
\begin{itemize}
    \item \( \alpha = 1 \): The function becomes linear; the output equals the input. In this case, the network is entirely linear.
    \item \( \alpha = 0 \): The function becomes the standard ReLU (hockey-stick function), which zeroes out all negative inputs. This introduces strong nonlinearity but can have issues with vanishing gradients\footnote{In this case the nonlinearity essentially partitions the space into regions where the output is linear from where it is zero. This allows the network to create local approximators.}.
    \item \( \alpha \in (0,1) \): Leaky ReLU allows a small gradient on the negative side, maintaining some responsiveness and avoiding the problem of vanishing gradients.
\end{itemize}

\begin{figure}[H]
    \centering
    \includegraphics[width=0.85\linewidth]{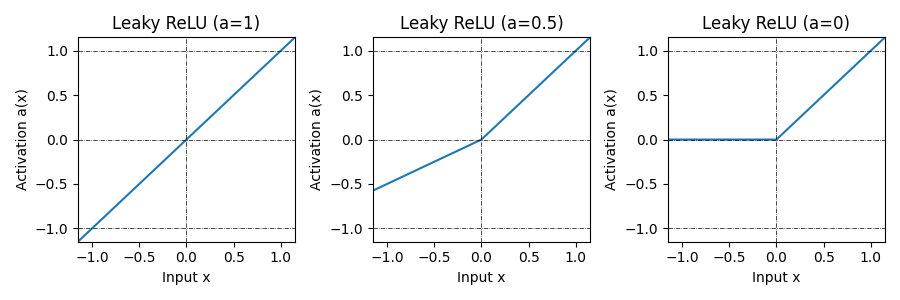}
    \caption{Leaky ReLU for different values of the hyperparameter \( \alpha \): linear (\( \alpha = 1 \)), semi-linear (\( \alpha = 0.5 \)), and ReLU (\( \alpha = 0 \)).}
    \label{fig:leaky_relu}
\end{figure}

When the activation function is linear ($\alpha=1$), the model itself becomes entirely linear. If all coefficients are left unrestricted, the number of layers becomes irrelevant: models with zero, one, two, or more layers will be observationally equivalent, and all reduce to standard linear quantile regression. However, in what follows, we will introduce regularization by placing a penalty on the parameters. In this case, while the model remains linear when ($\alpha=1$),  different numbers of layers will generate different regularization structures, because the penalty depends on how the parameters are distributed across layers. As a result, models with more layers will shrink differently, even if they remain linear in functional form.


\noindent
Different from the Leaky ReLU, this activation function will saturate the output for both large negative inputs and large positive ones. It also has the advantage of non-vanishing gradients.

\subsubsection{Penalization}

Since the models we consider are very densely parameterized, we add a penalty to regularize the problem. Even without any hidden layers—i.e., with only the output layer—we already have more parameters than predictors, which in our setting is a large number. This high dimensionality is typical in economic applications, where the number of predictors often rivals or exceeds the sample size. In such situations, complex models are prone to overfitting: they may fit the training data well but perform poorly out of sample, particularly in real-time forecasting. A widely adopted solution to this problem is to introduce a penalty that shrinks the individual parameters towards zero, thereby controlling model complexity and improving generalization. Regularization through penalization has a long history in mathematics and statistics \citep{Tikhonov1963,hoerl1970ridge}. It was introduced and studied in the context of forecasting with Big Data by \citet{DemolEtal2008}. For a general discussion, see \citet{DemolEtal2017}.

Specifically, we use a quadratic penalty, which corresponds to applying an \( \ell_2 \)-norm to all weights in the network:
\[
\text{Pen}(\theta, \psi, \lambda ) =  \lambda \left(  \sum_{i=1}^{d}\left(\|W^{(i)}\|_F^2 \right) + \|\gamma\|_2^2  \right)
\]

This regularization strategy is closely related to ridge regression. In fact, when the model includes only a single linear layer and no hidden layers, this formulation reduces to a ridge-regularized quantile regression model.

Ridge regression is known to be intimately related to data compression and regularization through factor structures, as shown in \citet{DemolEtal2008,DemolEtal2024}. Similarly, one could consider alternative penalties. Popular choices include sparsity-inducing penalties such as the \( \ell_1 \)-norm (Lasso), the \( \ell_0 \)-norm (subset selection), or combinations such as elastic nets or spike-and-slab priors. We choose to retain the \( \ell_2 \)-norm in our setup because there is neither strong theoretical motivation nor empirical evidence in favor of sparse forecasting models. In particular, \citet{GiannoneEtal2021} show that shrinkage---rather than sparsity---is the key to improving forecasting performance in high-dimensional settings.

More broadly, as in any forecasting model, there is a fundamental trade-off between bias and variance. The degree of penalization, governed by the hyperparameter \( \lambda \), determines where the model lies along this trade-off. At one extreme, when \( \lambda \rightarrow \infty \), all parameters are shrunk to zero, and the forecast collapses to the empirical unconditional quantile---resulting in zero variance but high bias, since all covariate information is discarded. At the other extreme, when \( \lambda = 0 \), there is no shrinkage, and the model fits both signal and noise---yielding low bias but potentially high variance. Intermediate values of \( \lambda \) balance this trade-off. In our implementation, the optimal degree of shrinkage is selected through out-of-sample validation, as discussed in the following section.

\subsubsection{Parameters and Hyperparameters}

In the discussion above, we implicitly distinguish between two sets of parameters:

\begin{itemize}

\item \( \theta \) are the model's learnable parameters, which are adjusted during training to minimize the loss function. Specifically, the weight matrices $[W^{(1)}, \ldots W^{(d)}] $ and biases $[b^{(1)}, \ldots b^{(d)}]$ of the $d$ layers of the deep neural network (DNN) and the coefficient of the output layer $\gamma, c$. This is usually a very large number of parameters.

\item The hyperparameters \( \psi \) are all other parameters that define the architecture, which are relatively few in number. They include the the number of layers $d$ in the neural network, the number of neurons $m_i, i = 1, \ldots, d$ in each layer, the choice of activation functions $a_i(\cdot;\alpha_i)$, for example, $a_i$ could be a Leaky ReLU with slope $\alpha_i$. Another hyperparameter controls  for the amount of penalization ( \( \lambda\) ), 

\end{itemize}

The distinction between parameters and hyperparameters is related to hierarchical Bayesian model \citep{GiannoneEtal2015,GiannoneEtal2021}. In our empirical analysis, hyperparameters are estimated or selected once based on out-of-sample performances over a validation sample and held fixed thereafter, while parameters, continuously re-estimated as new data becomes available (`online training'). This distinction is particularly important when we perform recursive estimation or online learning, as we do in our empirical implementation.

\section{Empirical Analysis}
\label{sec:empirical}

We begin by describing the data and how we split the sample; then we will discuss the estimation of parameters over a recursive training sample, hyperparameters over a validation sample, and finally evaluate model's performance over a test sample.

\subsection{Data}
We use the well-established dataset originally developed for macroeconomic forecasting of the US economy using a large number of predictors. The dataset was constructed by \citet{SW2002} when they introduced principal component regression in the context of Big Data. The dataset is a reference for all subsequent work on Big Data and machine learning, including \citet{ForniEtal2005} and \citet{DagostinoGiannone2012} with factor models, \citet{DemolEtal2008} and \citet{GiannoneEtal2021} with shrinkage, \citet{BanburaEtal2010} with large vector auto-regressions, \citet{Coulombe2022} and \citet{CarrieroEtal2024} with DNN.

The dataset consists of 130 predictors, including various monthly macroeconomic indicators, labor market variables, house prices, consumer and producer prices, money, credit, and asset prices. The sample ranges from February 1960 to January 2024, and all the data has been normalized.\footnote{The dataset has been updated by \citet{McCrackenNg2016} and is available through the Federal Reserve Economic Data (FRED).}

Most of the empirical analysis focuses on the one-month-ahead forecast of the change in the unemployment rate. To assess the robustness of the results, we also consider a longer horizon (one year ahead) and the one-month-ahead growth rate of industrial production.

\subsection{Estimation}
We conduct a simulated out-of-sample analysis by estimating the model recursively starting in January 1980. As detailed below, we split the out-of-sample period into two parts: we use the first part for selecting the architecture and the hyperparameters (validation), and the second part for testing the model's performance (testing).

\subsubsection{Sample Splitting}
To exploit the time series structure of the sample and use all available information, instead of splitting the data once into fixed training, validation, and test sets, we use an expanding training sample.

After each iteration, the window expands forward by one step, and the model is retrained on the new training window. For each step, we train the model on a data window from $1$ to $T$ and then test or validate the model on a future observation $T+h$. This expanding approach ensures the model continuously learns from past information. Alternatively, a rolling window with fixed size could be used.

Based on the trained predictions, we use a validation sample $T+h=T_1+1,\ldots,T_2$ to estimate the hyperparameters, and the rest of the sample $T+h=T_2+1,\ldots,T_3$ to evaluate the real-time model performance.

The validation and test samples are defined in terms of the forecast date $T+h$. As a result, the corresponding end of the estimation sample $T$ changes with the horizon. For the baseline case $h=1$, we set $T_1$ as January 1980, $T_2$ as January 2000, and $T_3$ as December 2024. The first forecast we evaluate is made in January 1980 ($T=T_1$) to predict February 1980 ($T_1+1$). The last forecast is made in November 1999 ($T=T_2-1$) to predict December 1999 ($T_2$).

In the robustness analysis, we also consider $h=12$. In that case, the first forecast is made in January 1980 to predict January 1981 ($T_1+12$). The last forecast is made in December 2023 ($T=T_3-12$) to predict December 2024 ($T_3$).

\subsubsection{Training: Estimation of \(\hat{\theta}\) for given  Architecture \((\psi\)) and Shrinkage (\(\lambda\)) and}

We approximate the quantiles of the predictive density as follows:
\[
Q_{Y_{T+h} \mid X_T = x_T}(\tau) \approx f_{\psi,\tau}(x_T; \theta).
\]

\noindent
We estimate the parameters \(\theta\) for a given set of hyperparameters \(\lambda\) over recursive samples:
\[
\hat \theta^{(T)}_{\lambda,\psi} = \arg \min_{\theta} \sum_{t+h=h+1}^{T} L_{\tau}(y_{t+h}, f_{\psi,\tau}(x_t; \theta)) + \text{Pen}(\theta, \psi, \lambda)
\]

\noindent
The out-of-sample predictions are:
\[
\hat Q^{( \psi,\lambda)}_{Y_{T+h} \mid X_T = x_T}(\tau) = f_{\psi,\tau}(x_T; \hat \theta^{(T)}_{\lambda,\psi})
\]

Out-of-sample means that the estimation of \(\theta\) does not use observations \(t > T\). Since the number of parameters \(\theta\) is large, estimation is prone to overfitting. We mitigate this with limited architectural complexity and regularization. After each training step, the model is evaluated on the next point in time. This recursive estimation preserves the temporal structure of the data and ensures that no future information is used when training the model. In the test period, hyperparameters are fixed, and parameters are updated recursively.

To save computation time, we initialize each new optimization using the estimates from the previous step: e.g., \(\hat \theta^{(T+1)}_{\lambda}\) is initialized with \(\hat \theta^{(T)}_{\lambda}\). In machine learning terminology, this is called warm-starting from a checkpoint. In the initial iteration, we use 500 epochs; in subsequent steps, we reduce to 100 epochs. The expanding window ensures that the model evolves gradually, helping maintaining prediction quality despite the warm-starting strategy.

We train the models using four Nvidia Quadro RTX 8000 GPUs and the PyTorch library for architecture design and gradient descent. Stochastic gradient descent produces locally optimal solutions, but due to its randomness, results may vary slightly depending on the seed. We tested sensitivity to different random seeds and found the results to be robust. 

\subsubsection{Validation and Testing: Selecting Architecture (\(\psi\)) and Shrinkage(\(\lambda\))}

Out-of-sample is used throughout. We split into validation, where we optimize hyperparameters and test where we freeze those hyperparameters and evaluate completely ut-of-sample.

As discussed above, the predictions
\[
\hat Q^{( \psi,\lambda)}_{Y_{T+h} \mid X_T = x_T}(\tau)=
f_{\psi,\tau}\left(x_T; \hat \theta^{(T)}_{\lambda,\psi} \right)
\]
still depend on the hyperparameters. We estimate these in the validation sample by minimizing the out-of-sample prediction loss:
\[
\hat \lambda,\hat \psi = \arg \min_{\lambda,\psi } \sum_{T+h=T_1+h}^{T_2} L_{\tau}\left(y_{T+h}, \hat Q^{( \psi,\lambda)}_{Y_{T+h} \mid X_T = x_T}(\tau) \right)
\]

We use the same loss function as in the training sample—the pinball loss—ensuring that the evaluation criterion aligns with the estimation target. This is consistent with the guidance in \citet{GiacominiKomunjer2005}, who emphasize that out-of-sample evaluation of quantile forecasts should be conducted using the same tick loss used for estimation. While their focus is on comparing forecast accuracy across models or relative to a benchmark, the same principle applies to selecting hyperparameters and architectures.

Below is the set of hyperparameters we have trained in the validation period (Table~\ref{tab:hyperparams}). When the number of nonlinear layers is zero, the model reduces to a linear regression, and there is no need to choose hidden dimensions or activation functions.\footnote{Therefore, the number of validation jobs is approximately $
3\times3\times 4\times 40 \times 8 = 11520
$.}

\begin{table}[!ht]
\centering
\begin{tabular}{ll}
\toprule
\textbf{Hyperparameter} & \textbf{Choices} \\
\midrule
\# of nonlinear layers & 0, 1, 2 \\
hidden dimension & 2, 4, 8 \\
activation & LeakyReLU(0.0), LeakyReLU(0.5), LeakyReLU(1.0) \\
L2 penalty & 40 choices from 0.2 to 10 (inclusive) in a logarithmic scale \\
loss function & mse, q0.05, q0.10, q0.25, q0.50, q0.75, q0.90, q0.95\\
\bottomrule
\end{tabular}
\caption{Hyperparameter grid explored during the validation stage. Note that the network architecture \(\psi\) is defined by the number of nonlinear layers, the hidden-layer width, and the activation function.}
\label{tab:hyperparams}
\end{table}

The performances in the validation sample are not genuine out-of-sample forecasts: they are subject to a look-ahead bias since the hyperparameters were selected to optimize performance over this very same evaluation period. To obtain an honest assessment of real-time performance, one must fix the hyperparameters based on the validation sample and then evaluate the model in a new test sample—never seen by the econometrician during either parameter or hyperparameter estimation. For this reason we evaluate model performance over the test sample \( T = T_2+1, \ldots, T_3 \), using the hyperparameters \(\hat{\lambda}, \hat{\psi}\) selected in the validation period. At each forecast time \(T\), parameters \(\theta_{\hat{\lambda}, \hat{\psi}}\) are re-estimated using data up to \(T\).

For \(T > T_2\), these predictions are fully real-time:
\[
\hat Q_{Y_{T+h} \mid X_T = x_T}(\tau)= g_{\hat \psi,\tau}\left(x_T; \hat \theta^{(T)}_{\hat \lambda, \hat \psi}\right)
\]


\noindent and the accuracy will be measured as
\[
\hat \lambda,\hat \psi = \arg \min_{\lambda,\psi } \sum_{T+h=T_2+h}^{T_3} L_{\tau}\left(y_{T+h}, \hat Q^{( \psi,\lambda)}_{Y_{T+h} \mid X_T = x_T}(\tau) \right)
\]

The difference in accuracy between validation and testing is informative and usually small if the hyperparameter space is not too large and if there are no major structural changes in the data.

Summing up, we split the out-of-sample in two parts, a validation and a test. We use the validation sample to optimize the hyperparameters, namely the architecture $\psi$ and the shrinkage parameter $\lambda$. Summarizing the outcome of this optimization directly is challenging because the hyperparameters involve multiple dimensions that interact in complex ways. For example, deeper networks with more neurons typically require stronger penalization to avoid overfitting.  In the following section we summarize all models in a single dimension that represent complexity.

\subsubsection{Complexity index and hyperparameter selection}\label{subsec:complexity}

A simple way to understand model complexity is through the in-sample variance of the forecasts, following the insights of \citet{GiannoneEtal2015}. Intuitively, more complex models fit more closely to the idiosyncrasies of the training data and therefore produce forecasts that fluctuate more in sample. Conversely, heavily penalized or highly constrained models generate smoother forecasts with lower variance.  

A natural benchmark is the naïve model with full shrinkage, which produces constant forecasts equal to the unconditional quantiles. This corresponds to a simple linear regression on a constant and contains no information from the predictors. We use these naïve forecasts as the reference point against which more complex models are evaluated.  

To formalize this idea, for any architecture $\psi$ and penalty parameter $\lambda$, we generate a sequence of \textit{fitted forecasts} over the initial estimation window $t+h=h+1,\ldots,T_1$:  
\[
\widetilde Q^{\psi,\lambda}_{t+h}(\tau)=f_{\psi,\tau}\big(x_t;\,\hat\theta^{(T_1)}_{\psi,\lambda}\big),
\qquad t+h=h+1,\ldots,T_1,
\]  
where $\hat\theta^{(T_1)}_{\psi,\lambda}$ is estimated using all data available up to $T_1$. We use the tilde notation to emphasize that these are in-sample fitted forecasts, in contrast to the recursive forecasts $\widehat Q$ used in the validation and testing periods.  

The mean fitted forecast is  
\[
\bar Q^{(0)}_{\psi,\lambda} = \frac{1}{T_1-h}\sum_{t+h=h+1}^{T_1}\widetilde Q^{\psi,\lambda}_{t+h}(\tau),
\]  
and the \textit{forecast variance} is  
\[
\operatorname{Var}_0(\psi,\lambda) 
= \frac{1}{T_1-h}\sum_{t+h=h+1}^{T_1}\Big(\widetilde Q^{\psi,\lambda}_{t+h}(\tau)-\bar Q^{(0)}_{\psi,\lambda}\Big)^2.
\]  

For each architecture, the unpenalized model $(\lambda=0)$ delivers the maximum forecast variance. Under full shrinkage ($\lambda \to \infty$), all weights are set to zero and predictions collapse to the unconditional quantiles. In this case, fitted forecasts are the same across all architectures, so we drop the index $\psi$ from the notation. The variance of these flat forecasts is  
\[
\operatorname{Var}_0^{\text{flat}} 
= \frac{1}{T_1-h}\sum_{t+h=h+1}^{T_1}\Big(\widetilde Q^{\text{naive}}_{t+h}(\tau)-\bar Q^{\text{naive}}\Big)^2,
\qquad 
\bar Q^{\text{naive}}=\frac{1}{T_1-h}\sum_{t+h=h+1}^{T_1}\widetilde Q^{\text{naive}}_{t+h}(\tau).
\]  

We then define the \textit{complexity index}  
\[
r(\psi,\lambda) = \frac{\operatorname{Var}_0(\psi,\lambda)}{\operatorname{Var}_0^{\max}} \;\;\in [0,1],
\]  
where $\operatorname{Var}_0^{\max}=\operatorname{Var}_0(\psi,0)$ is the maximum variance achieved by the unpenalized model. Since the denominator corresponds to the variance of the unconditional quantile forecasts, it is the same for all $\psi$, and we therefore omit the architecture index in the notation.  

In this definition, $r=0$ corresponds to flat, constant forecasts and the largest $r$ corresponds to the most complex, typically the most volatile, forecasts achievable with the model considered. Given the large set of predictors, the maximum will be close to 1, meaning that in sample it is possible to achieve an almost perfect fit.  

The complexity parameter $r$ provides a natural one-dimensional summary of the bias--variance trade-off. Models with low complexity ($r$ near zero) are highly shrunk and produce very smooth forecasts. In this region, variance is low but bias is high, since useful predictive information from the data is largely discarded. By contrast, models with high complexity ($r$ near one) produce forecasts that closely track the training sample. These forecasts have low bias, because they use the predictors aggressively, but high variance, because they are sensitive to noise and prone to overfitting. Optimal forecast performance is typically achieved at intermediate values of $r$, where the model captures the systematic signal in the data without being dominated by idiosyncratic noise.  

To operationalize this construction, we define a discrete grid of target complexity values,  
\[
\mathcal R = \{0.0, 0.1, 0.2, \ldots, 1.0\}.
\]  
For each $r \in \mathcal R$, we search over a set $S$ of candidate hyperparameters consisting of architectures $\psi$ and penalties $\lambda \in \{0\} \cup \{10^{-3},10^{-2.9},\ldots,10^2\}$ on a logarithmic scale. For each pair $(\psi,\lambda)$, we compute the implied complexity $r(\psi,\lambda)$. We then select the combination that produces a complexity index closest to the grid point:  
\[
(\psi_r,\lambda_r) \in 
\arg\min_{(\psi,\lambda)\in S} \big|\, r - r(\psi,\lambda)\,\big|.
\]  

This mapping compresses the high-dimensional hyperparameter space into a single, interpretable index. Each target value $r$ on the grid corresponds to a representative model $(\psi_r,\lambda_r)$, and performance can be evaluated as a function of $r$ rather than of individual hyperparameters. In this sense, our approach generalizes the linear case studied by \citet{DemolEtal2008} and \citet{BanburaEtal2010}, where shrinkage is the only dimension of regularization and the mapping between $\lambda$ and forecast variance is one-to-one. In the nonlinear setting considered here, by contrast, multiple combinations of architectures and shrinkage values can achieve a similar complexity index. To avoid reporting all possible combinations, we select one representative pair $(\psi_r,\lambda_r)$ for each grid point and summarize results only in terms of the effective complexity index $r$. The selected $(\psi_r,\lambda_r)$ should be seen as one representative among the set of models that yield the same complexity. The simplifying assumption is that models with similar complexity also display similar predictive performance, so that selecting hyperparameters in terms of $(\psi_r,\lambda_r)$ is approximately equivalent to selecting directly across $r$. Extensive checks over the full hyperparameter space confirm that this is a reasonable approximation, and that little is lost by summarizing performance in terms of the complexity index alone.  

Formally, the model used for forecasting is based on the triplet $(\hat r,\hat\lambda_{\hat r},\hat\psi_{\hat r})$. The complexity level $\hat r$ is selected in the validation sample by minimizing the quantile loss,  
\[
\hat r = \arg\min_{r } \sum_{T+h=T_1+h}^{T_2} L_{\tau}\left(y_{T+h}, \widehat Q^{( \psi_r,\lambda_r)}_{Y_{T+h} \mid X_T = x_T}(\tau) \right),
\]  
where $L_\tau$ denotes the quantile loss function. For each candidate value of $r$, the pair $(\hat\lambda_r,\hat\psi_r)$ is chosen among the set of hyperparameters that achieve the in-sample complexity closest to $r$. Thus the final choice $(\hat r,\hat\lambda_{\hat r},\hat\psi_{\hat r})$ should be interpreted as representative of this set. Empirical results based on the simplified procedure are very similar to those obtained from the full minimization over $(\lambda,\psi)$ discussed in the previous section,  
\[
\hat \lambda,\hat \psi = \arg \min_{\lambda,\psi } \sum_{T+h=T_1+h}^{T_2} L_{\tau}\left(y_{T+h}, \widehat Q^{( \psi,\lambda)}_{Y_{T+h} \mid X_T = x_T}(\tau) \right).
\]  
This empirical similarity allows us to present the results as a function of $r$ alone, rather than in terms of individual hyperparameters.\footnote{This equivalence between the simplified procedure based on $r$ and the full minimization over $(\lambda,\psi)$ is an empirical feature of our data and model. In other contexts, or with different architectures and datasets, the correspondence might be weaker, and the full hyperparameter search could be necessary.}  

\section{Results}
\label{sec:results}
We start our analysis by focusing on one-month-ahead forecasts of the unemployment rate. Later we will discuss robustness for other variables and horizons.  

In Table \ref{tab:Table2}, we report the average loss over the evaluation sample for each level of complexity, which we expressed in terms of the implied variance ratio. A variance ratio of zero implies full shrinkage (constant prediction), while a ratio of one corresponds to no shrinkage and maximum complexity. Losses are reported relative to the unconditional quantile benchmark, so the top row equals one by construction.


\begin{table}[htbp]
\centering
\tiny
\setlength{\tabcolsep}{1.0pt}
\begin{minipage}[t]{0.48\textwidth}
\centering
\begin{tabular}{>{\centering\arraybackslash}m{8mm} c|c|c|c|c|c|c|c}
\toprule
\multicolumn{9}{c}{\textbf{VALIDATION SAMPLE}} \\
\midrule
\multicolumn{2}{c}{} & \multicolumn{7}{c}{\textbf{Quantiles ($\tau$)}} \\
\cmidrule(lr){3-9}
\multicolumn{2}{c}{} & \textbf{0.05} & \textbf{0.10} & \textbf{0.25} & \textbf{0.50} & \textbf{0.75} & \textbf{0.90} & \textbf{0.95} \\
\midrule
\multirow{11}{8mm}{\centering\rotatebox[origin=c]{90}{\textbf{Complexity}}}
& 0.0 & \cellcolor{baseline} 95.20 & \cellcolor{baseline} 81.57 & \cellcolor{baseline} 65.62 & \cellcolor{baseline} 60.82 & \cellcolor{baseline} 70.29 & \cellcolor{baseline} 89.41 & \cellcolor{baseline} 107.81 \\
& 0.1 & \cellcolor{greenlight} \begin{tabular}[c]{@{}c@{}} -1.47 \\ (7.35) \end{tabular} & \cellcolor{greenmedium} \begin{tabular}[c]{@{}c@{}} -4.76 \\ (3.42) \end{tabular} & \cellcolor{greenmedium} \begin{tabular}[c]{@{}c@{}} -3.58 \\ (2.69) \end{tabular} & \cellcolor{greenlight} \begin{tabular}[c]{@{}c@{}} -3.43 \\ (2.68) \end{tabular} & \cellcolor{greendark} \begin{tabular}[c]{@{}c@{}} -8.33 \\ (3.33) \end{tabular} & \cellcolor{greendark} \begin{tabular}[c]{@{}c@{}} -15.10 \\ (3.92) \end{tabular} & \cellcolor{greendark} \begin{tabular}[c]{@{}c@{}} -21.44 \\ (7.56) \end{tabular} \\
& 0.2 & \cellcolor{greenlight} \begin{tabular}[c]{@{}c@{}} -0.55 \\ (8.18) \end{tabular} & \cellcolor{greenlight} \begin{tabular}[c]{@{}c@{}} -2.90 \\ (4.06) \end{tabular} & \cellcolor{greenlight} \begin{tabular}[c]{@{}c@{}} -2.77 \\ (3.60) \end{tabular} & \cellcolor{greenlight} \begin{tabular}[c]{@{}c@{}} -2.15 \\ (2.87) \end{tabular} & \cellcolor{greendark} \begin{tabular}[c]{@{}c@{}} -9.97 \\ (3.74) \end{tabular} & \cellcolor{greendark} \begin{tabular}[c]{@{}c@{}} -15.10 \\ (3.92) \end{tabular} & \cellcolor{greendark} \begin{tabular}[c]{@{}c@{}} -21.13 \\ (11.34) \end{tabular} \\
& 0.3 & \cellcolor{redlight} \begin{tabular}[c]{@{}c@{}} 0.51 \\ (8.32) \end{tabular} & \cellcolor{greenlight} \begin{tabular}[c]{@{}c@{}} -1.57 \\ (4.52) \end{tabular} & \cellcolor{greenlight} \begin{tabular}[c]{@{}c@{}} -2.59 \\ (3.98) \end{tabular} & \cellcolor{greenlight} \begin{tabular}[c]{@{}c@{}} -0.92 \\ (3.16) \end{tabular} & \cellcolor{greendark} \begin{tabular}[c]{@{}c@{}} -10.68 \\ (3.88) \end{tabular} & \cellcolor{greendark} \begin{tabular}[c]{@{}c@{}} -17.41 \\ (6.30) \end{tabular} & \cellcolor{greenlight} \begin{tabular}[c]{@{}c@{}} -13.93 \\ (13.95) \end{tabular} \\
& 0.4 & \cellcolor{redlight} \begin{tabular}[c]{@{}c@{}} 7.96 \\ (8.51) \end{tabular} & \cellcolor{redlight} \begin{tabular}[c]{@{}c@{}} 0.59 \\ (4.69) \end{tabular} & \cellcolor{redlight} \begin{tabular}[c]{@{}c@{}} 0.01 \\ (3.98) \end{tabular} & \cellcolor{redlight} \begin{tabular}[c]{@{}c@{}} 3.12 \\ (3.68) \end{tabular} & \cellcolor{greendark} \begin{tabular}[c]{@{}c@{}} -7.79 \\ (4.32) \end{tabular} & \cellcolor{greenlight} \begin{tabular}[c]{@{}c@{}} -6.58 \\ (7.53) \end{tabular} & \cellcolor{greenlight} \begin{tabular}[c]{@{}c@{}} -3.06 \\ (15.84) \end{tabular} \\
& 0.5 & \cellcolor{reddark} \begin{tabular}[c]{@{}c@{}} 24.86 \\ (11.05) \end{tabular} & \cellcolor{redlight} \begin{tabular}[c]{@{}c@{}} 4.29 \\ (5.06) \end{tabular} & \cellcolor{redmedium} \begin{tabular}[c]{@{}c@{}} 5.37 \\ (3.96) \end{tabular} & \cellcolor{redmedium} \begin{tabular}[c]{@{}c@{}} 5.76 \\ (3.94) \end{tabular} & \cellcolor{greenlight} \begin{tabular}[c]{@{}c@{}} -3.91 \\ (4.95) \end{tabular} & \cellcolor{greenlight} \begin{tabular}[c]{@{}c@{}} -3.19 \\ (7.87) \end{tabular} & \cellcolor{redlight} \begin{tabular}[c]{@{}c@{}} 14.34 \\ (17.21) \end{tabular} \\
& 0.6 & \cellcolor{reddark} \begin{tabular}[c]{@{}c@{}} 32.87 \\ (12.38) \end{tabular} & \cellcolor{reddark} \begin{tabular}[c]{@{}c@{}} 10.15 \\ (5.60) \end{tabular} & \cellcolor{redmedium} \begin{tabular}[c]{@{}c@{}} 6.32 \\ (4.16) \end{tabular} & \cellcolor{reddark} \begin{tabular}[c]{@{}c@{}} 8.39 \\ (4.07) \end{tabular} & \cellcolor{redlight} \begin{tabular}[c]{@{}c@{}} 3.41 \\ (5.76) \end{tabular} & \cellcolor{redlight} \begin{tabular}[c]{@{}c@{}} 6.52 \\ (9.56) \end{tabular} & \cellcolor{redlight} \begin{tabular}[c]{@{}c@{}} 18.22 \\ (17.43) \end{tabular} \\
& 0.7 & \cellcolor{reddark} \begin{tabular}[c]{@{}c@{}} 33.49 \\ (12.96) \end{tabular} & \cellcolor{reddark} \begin{tabular}[c]{@{}c@{}} 36.27 \\ (8.44) \end{tabular} & \cellcolor{reddark} \begin{tabular}[c]{@{}c@{}} 11.05 \\ (4.83) \end{tabular} & \cellcolor{reddark} \begin{tabular}[c]{@{}c@{}} 9.42 \\ (4.29) \end{tabular} & \cellcolor{redlight} \begin{tabular}[c]{@{}c@{}} 0.15 \\ (5.26) \end{tabular} & \cellcolor{redlight} \begin{tabular}[c]{@{}c@{}} 9.77 \\ (9.49) \end{tabular} & \cellcolor{reddark} \begin{tabular}[c]{@{}c@{}} 31.04 \\ (18.78) \end{tabular} \\
& 0.8 & \cellcolor{reddark} \begin{tabular}[c]{@{}c@{}} 34.57 \\ (14.05) \end{tabular} & \cellcolor{reddark} \begin{tabular}[c]{@{}c@{}} 37.94 \\ (8.69) \end{tabular} & \cellcolor{reddark} \begin{tabular}[c]{@{}c@{}} 18.53 \\ (5.24) \end{tabular} & \cellcolor{reddark} \begin{tabular}[c]{@{}c@{}} 10.56 \\ (4.46) \end{tabular} & \cellcolor{redlight} \begin{tabular}[c]{@{}c@{}} 6.39 \\ (6.05) \end{tabular} & \cellcolor{redmedium} \begin{tabular}[c]{@{}c@{}} 18.02 \\ (11.49) \end{tabular} & \cellcolor{reddark} \begin{tabular}[c]{@{}c@{}} 41.85 \\ (19.62) \end{tabular} \\
& 0.9 & \cellcolor{reddark} \begin{tabular}[c]{@{}c@{}} 36.19 \\ (14.21) \end{tabular} & \cellcolor{reddark} \begin{tabular}[c]{@{}c@{}} 39.35 \\ (8.07) \end{tabular} & \cellcolor{reddark} \begin{tabular}[c]{@{}c@{}} 16.61 \\ (5.05) \end{tabular} & \cellcolor{reddark} \begin{tabular}[c]{@{}c@{}} 13.58 \\ (4.64) \end{tabular} & \cellcolor{redlight} \begin{tabular}[c]{@{}c@{}} 5.52 \\ (5.82) \end{tabular} & \cellcolor{redlight} \begin{tabular}[c]{@{}c@{}} 11.28 \\ (10.48) \end{tabular} & \cellcolor{reddark} \begin{tabular}[c]{@{}c@{}} 36.76 \\ (19.38) \end{tabular} \\
& 1.0 & \cellcolor{reddark} \begin{tabular}[c]{@{}c@{}} 47.72 \\ (14.91) \end{tabular} & \cellcolor{reddark} \begin{tabular}[c]{@{}c@{}} 47.76 \\ (11.49) \end{tabular} & \cellcolor{reddark} \begin{tabular}[c]{@{}c@{}} 19.57 \\ (5.50) \end{tabular} & \cellcolor{reddark} \begin{tabular}[c]{@{}c@{}} 16.08 \\ (4.69) \end{tabular} & \cellcolor{redmedium} \begin{tabular}[c]{@{}c@{}} 8.06 \\ (5.87) \end{tabular} & \cellcolor{reddark} \begin{tabular}[c]{@{}c@{}} 20.06 \\ (11.72) \end{tabular} & \cellcolor{reddark} \begin{tabular}[c]{@{}c@{}} 39.13 \\ (20.77) \end{tabular} \\
\bottomrule
\end{tabular}
\end{minipage}%
\hfill
\begin{minipage}[t]{0.48\textwidth}
\centering
\begin{tabular}{>{\centering\arraybackslash}m{8mm} c|c|c|c|c|c|c|c}
\toprule
\multicolumn{9}{c}{\textbf{TEST SAMPLE}} \\
\midrule
\multicolumn{2}{c}{} & \multicolumn{7}{c}{\textbf{Quantiles ($\tau$)}} \\
\cmidrule(lr){3-9}
\multicolumn{2}{c}{} & \textbf{0.05} & \textbf{0.10} & \textbf{0.25} & \textbf{0.50} & \textbf{0.75} & \textbf{0.90} & \textbf{0.95} \\
\midrule
\multirow{11}{8mm}{\centering\rotatebox[origin=c]{90}{\textbf{Complexity}}}
& 0.0 & \cellcolor{baseline} 80.65 & \cellcolor{baseline} 63.69 & \cellcolor{baseline} 54.43 & \cellcolor{baseline} 54.04 & \cellcolor{baseline} 61.63 & \cellcolor{baseline} 79.56 & \cellcolor{baseline} 95.59 \\
& 0.1 & \cellcolor{greenlight} \begin{tabular}[c]{@{}c@{}} -0.10 \\ (0.09) \end{tabular} & \cellcolor{greenlight} \begin{tabular}[c]{@{}c@{}} -0.24 \\ (2.89) \end{tabular} & \cellcolor{greenlight} \begin{tabular}[c]{@{}c@{}} -0.99 \\ (1.83) \end{tabular} & \cellcolor{greenlight} \begin{tabular}[c]{@{}c@{}} -2.52 \\ (2.08) \end{tabular} & \cellcolor{greenlight} \begin{tabular}[c]{@{}c@{}} -0.05 \\ (0.48) \end{tabular} & \cellcolor{greenmedium} \begin{tabular}[c]{@{}c@{}} -9.10 \\ (6.20) \end{tabular} & \cellcolor{greendark} \begin{tabular}[c]{@{}c@{}} -16.59 \\ (7.76) \end{tabular} \\
& 0.2 & \cellcolor{greendark} \begin{tabular}[c]{@{}c@{}} -4.50 \\ (1.25) \end{tabular} & \cellcolor{redlight} \begin{tabular}[c]{@{}c@{}} 1.65 \\ (3.69) \end{tabular} & \cellcolor{redlight} \begin{tabular}[c]{@{}c@{}} 0.26 \\ (2.83) \end{tabular} & \cellcolor{greenlight} \begin{tabular}[c]{@{}c@{}} -3.16 \\ (2.54) \end{tabular} & \cellcolor{greenlight} \begin{tabular}[c]{@{}c@{}} -6.46 \\ (5.38) \end{tabular} & \cellcolor{greenmedium} \begin{tabular}[c]{@{}c@{}} -9.10 \\ (6.20) \end{tabular} & \cellcolor{greendark} \begin{tabular}[c]{@{}c@{}} -27.46 \\ (12.30) \end{tabular} \\
& 0.3 & \cellcolor{greenlight} \begin{tabular}[c]{@{}c@{}} -1.62 \\ (6.29) \end{tabular} & \cellcolor{redlight} \begin{tabular}[c]{@{}c@{}} 3.33 \\ (4.11) \end{tabular} & \cellcolor{redlight} \begin{tabular}[c]{@{}c@{}} 2.23 \\ (3.46) \end{tabular} & \cellcolor{greenlight} \begin{tabular}[c]{@{}c@{}} -1.87 \\ (3.09) \end{tabular} & \cellcolor{greenmedium} \begin{tabular}[c]{@{}c@{}} -7.26 \\ (5.59) \end{tabular} & \cellcolor{greendark} \begin{tabular}[c]{@{}c@{}} -19.94 \\ (10.71) \end{tabular} & \cellcolor{greendark} \begin{tabular}[c]{@{}c@{}} -27.19 \\ (12.02) \end{tabular} \\
& 0.4 & \cellcolor{redlight} \begin{tabular}[c]{@{}c@{}} 7.75 \\ (9.83) \end{tabular} & \cellcolor{redlight} \begin{tabular}[c]{@{}c@{}} 5.13 \\ (4.27) \end{tabular} & \cellcolor{redlight} \begin{tabular}[c]{@{}c@{}} 4.82 \\ (4.00) \end{tabular} & \cellcolor{redlight} \begin{tabular}[c]{@{}c@{}} 0.07 \\ (3.96) \end{tabular} & \cellcolor{greenmedium} \begin{tabular}[c]{@{}c@{}} -8.17 \\ (6.13) \end{tabular} & \cellcolor{greendark} \begin{tabular}[c]{@{}c@{}} -19.50 \\ (10.72) \end{tabular} & \cellcolor{greendark} \begin{tabular}[c]{@{}c@{}} -24.05 \\ (11.52) \end{tabular} \\
& 0.5 & \cellcolor{redmedium} \begin{tabular}[c]{@{}c@{}} 14.43 \\ (9.60) \end{tabular} & \cellcolor{reddark} \begin{tabular}[c]{@{}c@{}} 10.43 \\ (4.76) \end{tabular} & \cellcolor{reddark} \begin{tabular}[c]{@{}c@{}} 8.32 \\ (4.17) \end{tabular} & \cellcolor{redlight} \begin{tabular}[c]{@{}c@{}} 1.55 \\ (3.88) \end{tabular} & \cellcolor{greenlight} \begin{tabular}[c]{@{}c@{}} -5.43 \\ (6.46) \end{tabular} & \cellcolor{greenmedium} \begin{tabular}[c]{@{}c@{}} -17.57 \\ (10.98) \end{tabular} & \cellcolor{greendark} \begin{tabular}[c]{@{}c@{}} -19.63 \\ (11.20) \end{tabular} \\
& 0.6 & \cellcolor{reddark} \begin{tabular}[c]{@{}c@{}} 21.15 \\ (10.16) \end{tabular} & \cellcolor{reddark} \begin{tabular}[c]{@{}c@{}} 14.77 \\ (5.29) \end{tabular} & \cellcolor{reddark} \begin{tabular}[c]{@{}c@{}} 9.86 \\ (4.24) \end{tabular} & \cellcolor{redlight} \begin{tabular}[c]{@{}c@{}} 3.49 \\ (3.12) \end{tabular} & \cellcolor{redlight} \begin{tabular}[c]{@{}c@{}} 0.24 \\ (5.82) \end{tabular} & \cellcolor{greenlight} \begin{tabular}[c]{@{}c@{}} -4.13 \\ (9.83) \end{tabular} & \cellcolor{greendark} \begin{tabular}[c]{@{}c@{}} -17.43 \\ (10.43) \end{tabular} \\
& 0.7 & \cellcolor{reddark} \begin{tabular}[c]{@{}c@{}} 30.01 \\ (11.35) \end{tabular} & \cellcolor{reddark} \begin{tabular}[c]{@{}c@{}} 28.15 \\ (6.55) \end{tabular} & \cellcolor{reddark} \begin{tabular}[c]{@{}c@{}} 15.62 \\ (4.02) \end{tabular} & \cellcolor{reddark} \begin{tabular}[c]{@{}c@{}} 6.48 \\ (3.16) \end{tabular} & \cellcolor{greenlight} \begin{tabular}[c]{@{}c@{}} -2.67 \\ (6.18) \end{tabular} & \cellcolor{greenlight} \begin{tabular}[c]{@{}c@{}} -5.90 \\ (9.33) \end{tabular} & \cellcolor{redlight} \begin{tabular}[c]{@{}c@{}} 5.07 \\ (11.10) \end{tabular} \\
& 0.8 & \cellcolor{reddark} \begin{tabular}[c]{@{}c@{}} 37.47 \\ (12.15) \end{tabular} & \cellcolor{reddark} \begin{tabular}[c]{@{}c@{}} 25.52 \\ (6.98) \end{tabular} & \cellcolor{reddark} \begin{tabular}[c]{@{}c@{}} 21.16 \\ (3.91) \end{tabular} & \cellcolor{reddark} \begin{tabular}[c]{@{}c@{}} 9.70 \\ (3.21) \end{tabular} & \cellcolor{redlight} \begin{tabular}[c]{@{}c@{}} 1.36 \\ (5.64) \end{tabular} & \cellcolor{redlight} \begin{tabular}[c]{@{}c@{}} 1.22 \\ (8.29) \end{tabular} & \cellcolor{redlight} \begin{tabular}[c]{@{}c@{}} 15.74 \\ (13.11) \end{tabular} \\
& 0.9 & \cellcolor{reddark} \begin{tabular}[c]{@{}c@{}} 39.29 \\ (12.66) \end{tabular} & \cellcolor{reddark} \begin{tabular}[c]{@{}c@{}} 23.92 \\ (5.77) \end{tabular} & \cellcolor{reddark} \begin{tabular}[c]{@{}c@{}} 21.23 \\ (4.08) \end{tabular} & \cellcolor{reddark} \begin{tabular}[c]{@{}c@{}} 17.74 \\ (3.48) \end{tabular} & \cellcolor{redlight} \begin{tabular}[c]{@{}c@{}} 0.79 \\ (5.52) \end{tabular} & \cellcolor{greenlight} \begin{tabular}[c]{@{}c@{}} -3.16 \\ (9.01) \end{tabular} & \cellcolor{redlight} \begin{tabular}[c]{@{}c@{}} 12.04 \\ (11.92) \end{tabular} \\
& 1.0 & \cellcolor{reddark} \begin{tabular}[c]{@{}c@{}} 66.45 \\ (16.78) \end{tabular} & \cellcolor{reddark} \begin{tabular}[c]{@{}c@{}} 24.08 \\ (6.66) \end{tabular} & \cellcolor{reddark} \begin{tabular}[c]{@{}c@{}} 23.27 \\ (4.79) \end{tabular} & \cellcolor{reddark} \begin{tabular}[c]{@{}c@{}} 27.20 \\ (4.94) \end{tabular} & \cellcolor{redlight} \begin{tabular}[c]{@{}c@{}} 2.53 \\ (5.78) \end{tabular} & \cellcolor{redlight} \begin{tabular}[c]{@{}c@{}} 4.98 \\ (8.31) \end{tabular} & \cellcolor{redmedium} \begin{tabular}[c]{@{}c@{}} 28.56 \\ (19.21) \end{tabular} \\
\bottomrule
\end{tabular}
\end{minipage}

\vspace{0.3cm}
\caption{Accuracy of the forecast of the change in the unemployment rate 1 month ahead. The first row shows absolute loss of naive model based on recursively estimated unconditional quantiles, normalized by $\tau(1-\tau)$. Subsequent rows show differences of model loss relative to naive baseline, normalized by $\tau(1-\tau)$. Negative values (green) indicate model outperforms naive; positive values (red) indicate underperformance. HAC standard errors in parentheses (robust to autocorrelation and heteroskedasticity), also normalized by $\tau(1-\tau)$. Color darkness based on $|\text{differential}/\text{standard error}|$ ratio using thresholds 1.28 and 1.65 (90th and 95th quantiles of the normal distribution): light shading below 1.28, medium between 1.28 and 1.65, darker if larger than 1.65.}
\label{tab:Table2}
\end{table}

This highlights the fundamental bias-variance trade-off in statistical learning. More complexity allows the model to capture more structure from the data but increases variance. More shrinkage stabilizes predictions but may lead to bias. The optimal point minimizes expected loss by balancing these effects.

For lower quantiles, the optimal complexity ratio around \(0.0{-}0.2\); for upper quantiles, it is closer to \(0.2{-}0.4\). The loss is lower for upper quantiles, suggesting greater predictability in the upper tail (i.e., rising unemployment). The need for more complexity in this region suggests that the predictors contain useful information. Conversely, for lower quantiles, optimal performance is achieved with limited complexity, suggesting limited information in the predictors for this region. In this case, variance reduction dominates.

For the upper quantiles, the model delivers clear gains relative to the unconditional quantile, thanks to its ability to extract predictive signals from the large information set through a flexible architecture. The more predictive content there is in the data, the broader the range of complexity ratios for which the model outperforms the naïve benchmark. However, even in cases of strong predictability, the danger of overfitting is always present. For instance, when the variance ratio exceeds 0.8, predictive losses increase sharply, and all forecasts deteriorate below the performance of the unconditional model. This underscores the critical role of out-of-sample validation in selecting the hyperparameters that govern model complexity.

Optimal complexity is fund to correspond to a range of variance ratios around 0.1 to 0.3 across quantiles.

We compare the performances in the test sample in order to study the stability of the performances between the validation and the testing sample. Comparing table for validation and testing we see very similar patters, in particular similar improvements over relevant ranges. This highlights the stability of the model performances across the two sample. Specifically, we see that in the test period the optimal complexity ratio is in a similar range (0.1 to 0.3)

\begin{figure}[htbp]
  \centering

  \begin{subfigure}[b]{0.6\textwidth}
    \centering
    \includegraphics[width=\linewidth]{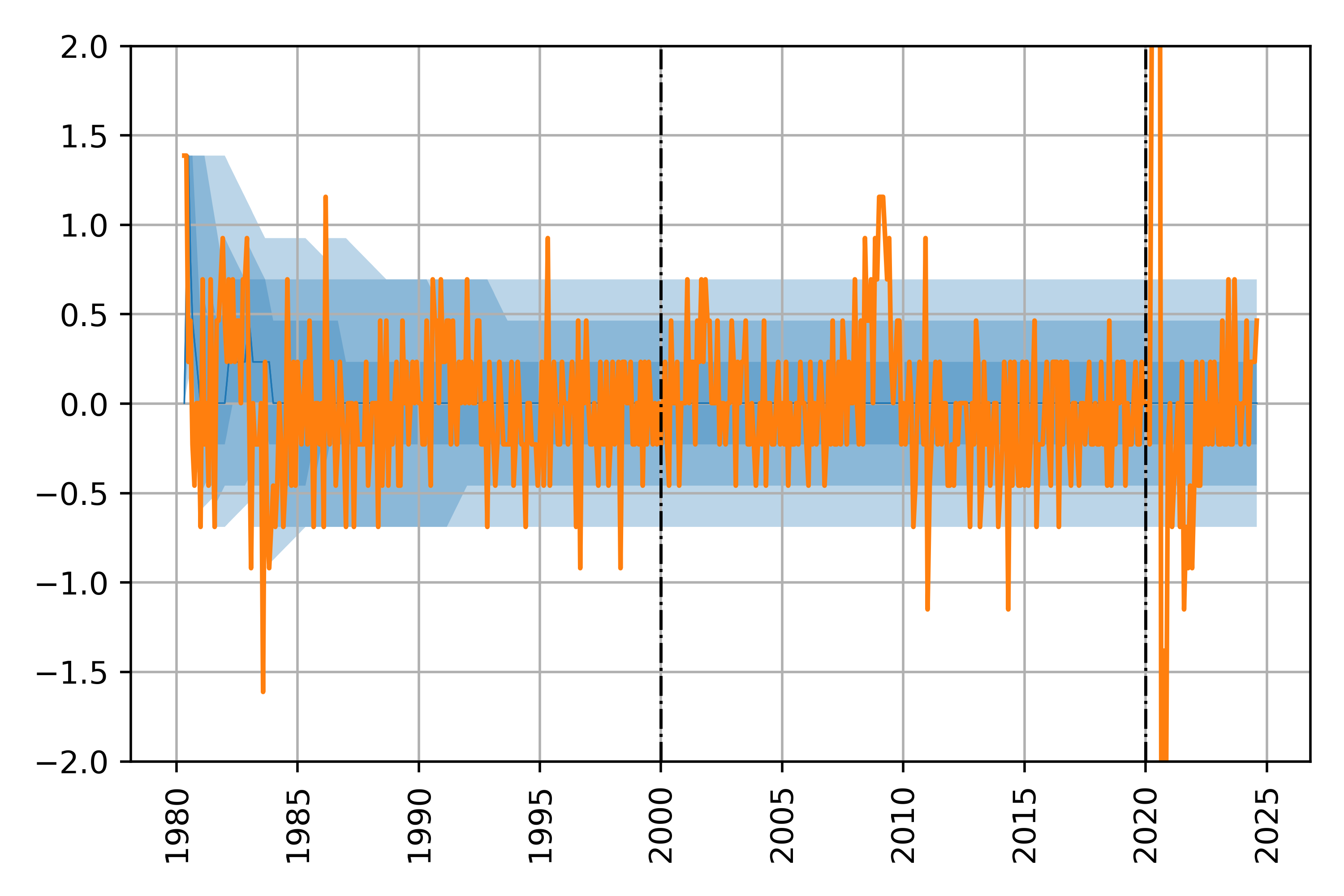}
    \caption{Full shrinkage}
    \label{fig:shrinkage-full}
  \end{subfigure}

  \begin{subfigure}[b]{0.6\textwidth}
    \centering
    \includegraphics[width=\linewidth]{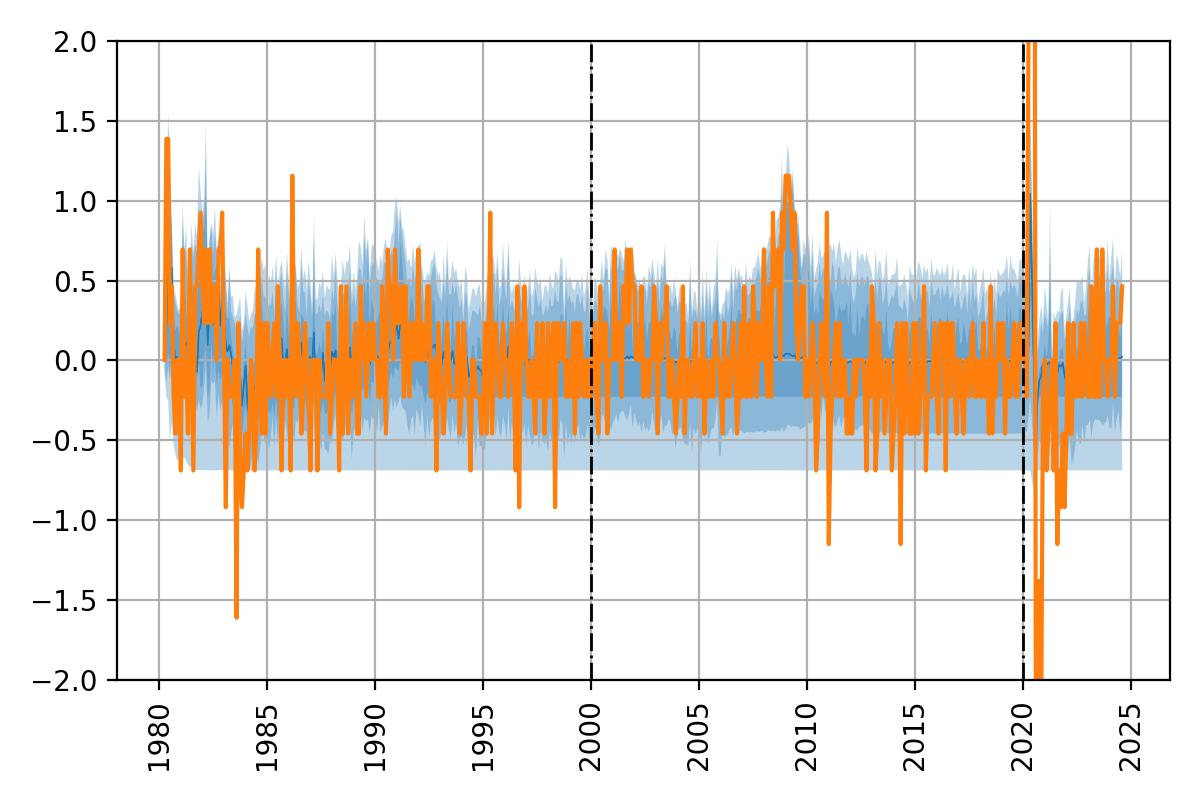}
    \caption{Validation optimal}
    \label{fig:shrinkage-opt}
  \end{subfigure}

  \begin{subfigure}[b]{0.6\textwidth}
    \centering
    \includegraphics[width=\linewidth]{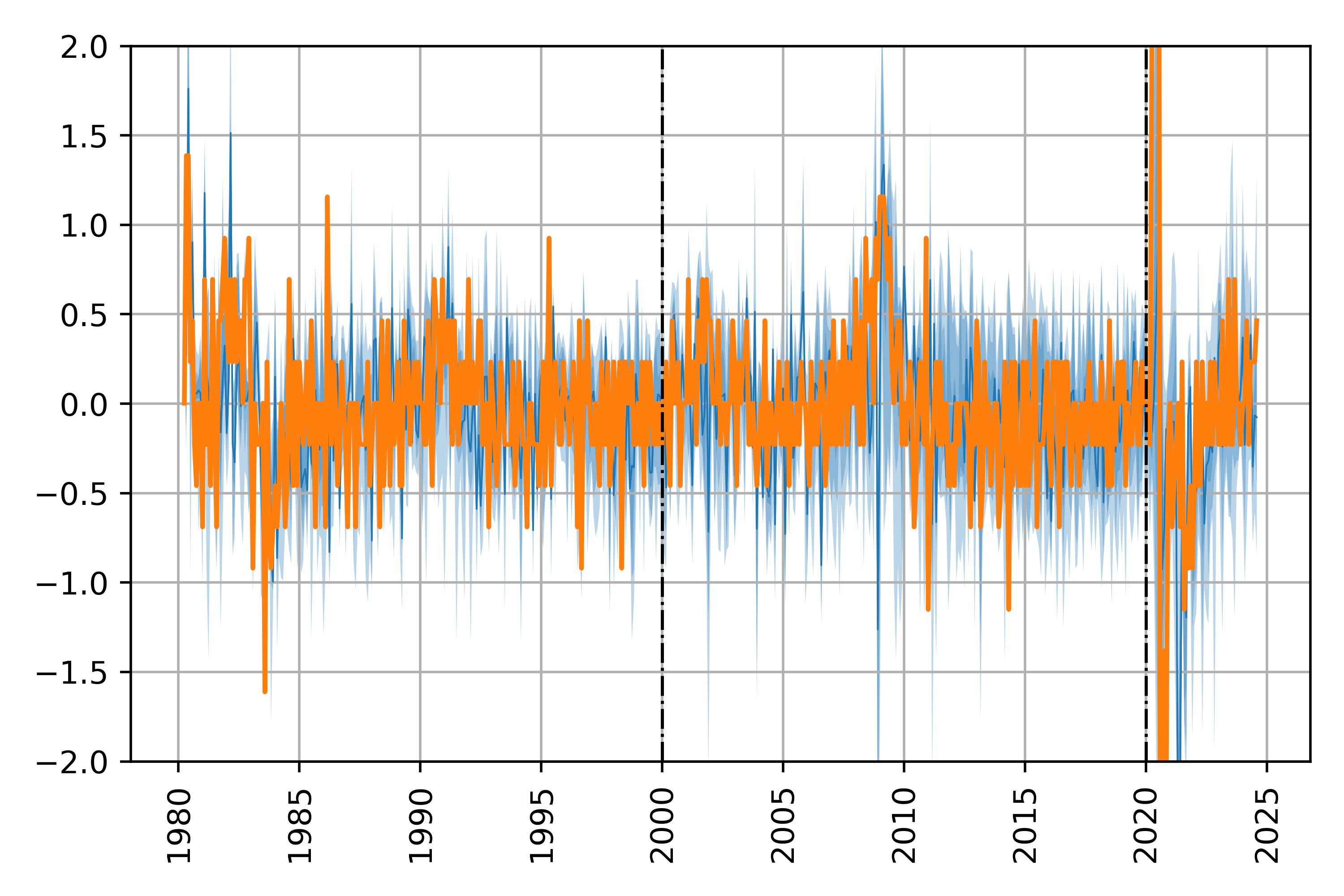}
    \caption{No shrinkage}
    \label{fig:shrinkage-none}
  \end{subfigure}

    \caption{Forecast of the change in the unemployment rate 1 month ahead. Actual (solid, orange) and quantiles of the predicted distribution (shaded, blue) with: (a) Full shrinkage (b) Optimal shrinkage (c) No shrinkage.}
  \label{fig:shrinkage}
\end{figure}

This is further clarified by looking at predictions over time. We report predicted quantiles in Figure~\ref{fig:shrinkage}, alongside the actual unemployment rate. The chart shows forecasts over the entire out-of-sample period, including both validation and testing. The vertical dotted line separates the two. Recall that forecasts in the testing period are \textit{fully real-time}, since parameter estimates are recursive and hyperparameters are held fixed at the values selected during the validation period.

The central panel shows forecasts based on the architecture and hyperparameters selected in the validation sample. Forecasts in the validation period are not truly out-of-sample because hyperparameters are optimized using that sample. Thus, some overfitting is possible. However, because the number of hyperparameters is small relative to the sample size, the risk is limited.

To illustrate the role of complexity, we also show two extremes. The left panel imposes infinite shrinkage (variance ratio = 0), collapsing predictions to the unconditional quantiles. The right panel applies no shrinkage (variance ratio = 1), resulting in excessively volatile forecasts, indicating severe overfitting.

These three panels clearly display the bias-variance trade-off. At one extreme, excessive shrinkage eliminates all information, yielding stable but biased forecasts. At the other extreme, the absence of restrictions allow to incorporates all information exploiting the richeness of the unconstraned model, yielding low bias but high variance. The middle panel strikes an optimal balance. Since validation is done out-of-sample, this trade-off is clearly observable.

The middle panel shows that the model captures rising downside risk when it emerges. It also reflects business cycle asymmetry: downside risks increase, while upside risks (lower quantiles of unemployment) remain relatively flat. Notably, the similarity between forecasts in validation and testing periods indicates that overfitting due to hyperparameter selection is limited. This highlights the value of using out-of-sample performance as the criterion for model selection---a core innovation of machine learning.

In summary, the model detects time-varying labor market risk, especially on the downside. It identifies emerging vulnerabilities using a large set of macro indicators.

Importantly, all forecasts are generated in a fully automated, hands-off-the-wheel fashion. No manual selection of predictors or model tuning is used. All decisions---architecture, shrinkage, and learning rate---are guided by out-of-sample validation. Remarkably, this approach reproduces key insights from earlier studies \citep[e.g.,][]{AdrianEtal2019,kiley,boyarchenko2023outlook}, but with no handcrafted inputs. Previous studies used two-step procedures with expert-selected factors; here, all features are learned from data.

Through careful regularization and model selection, the DNN handles a very flexible architecture without overfitting. We attribute this robustness to the combination of shrinkage and the DNN's ability to operate on low-dimensional manifolds, as discussed in \citet{GhigliazzaEtal2020}.

\subsection{Diving Deeper}

We will now study the role of nonlinearities, and then assess robustness by considering different horizons and target variables. Specifically, we examine one-month-ahead unemployment, twelve-month-ahead unemployment, and one-month-ahead industrial production growth. These exercises demonstrate the flexibility and robustness of quantile estimation via DNNs.

\subsubsection{The role of non-linearities}

In general, the model allows for complex non-linear interactions. However, one should check that a linear model is insufficient to match the problem's complexity, which, in this case is a penalized linear quantile regression, albeit over-parametrized.

\begin{figure}[H]
  \centering
  \begin{subfigure}[b]{0.45\textwidth}
    \centering
    \includegraphics[width=\linewidth]{figs/leakyrelu_validation_optimal.png}
    \caption{Leaky ReLU activation}
    \label{fig:effect-nonlin-a}
  \end{subfigure}\hfill
  \begin{subfigure}[b]{0.45\textwidth}
    \centering
    \includegraphics[width=\linewidth]{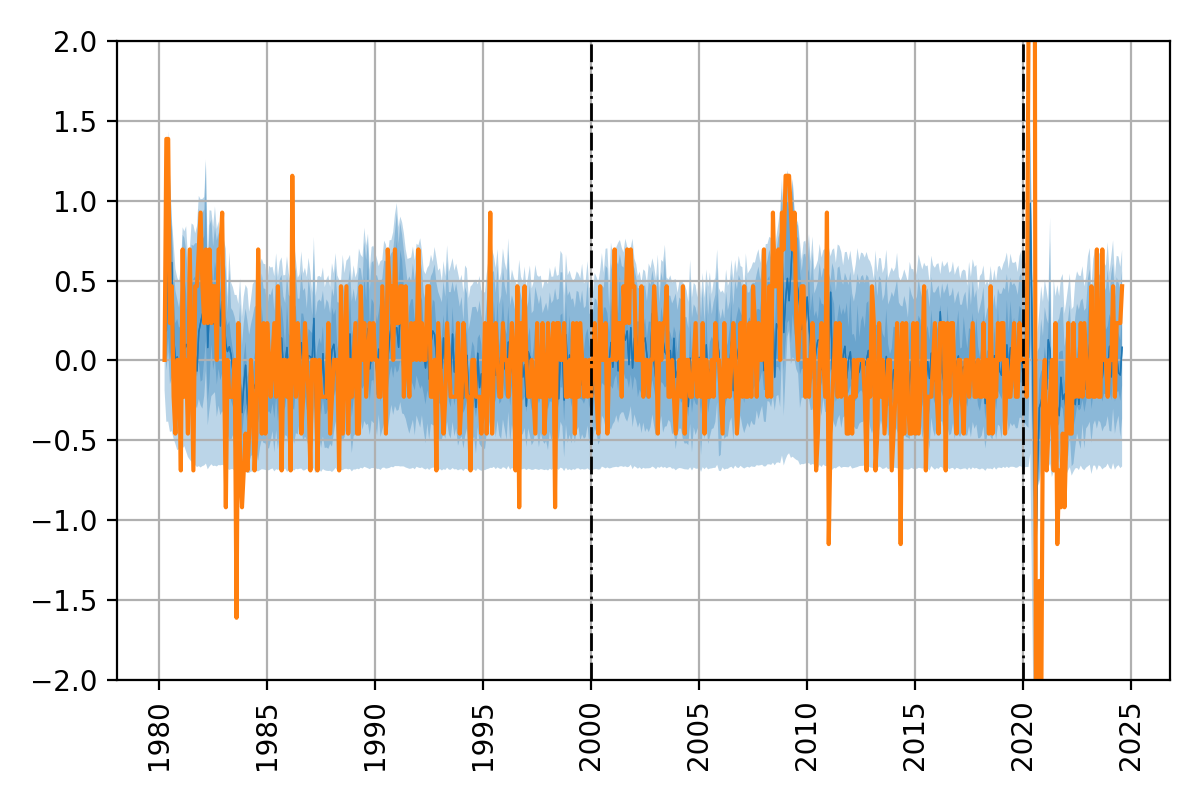}
    \caption{Linear activation}
    \label{fig:effect-nonlin-b}
  \end{subfigure}
    \caption{Forecast of the change in the unemployment rate 1 month ahead. Actual (solid, orange) and quantiles of the predicted distribution (shaded, blue) with:  (a) Leaky ReLU activation (b) Linear activation.}
  \label{fig:effect_nonlinearities}
\end{figure}

In the left panel of Figure~\ref{fig:effect_nonlinearities}, we report the predictions from the full model, where the slope of the Leaky ReLU is treated as a hyperparameter and chosen (along with others) to optimize performance in the validation sample. This is the same model as the one shown in the middle panel of Figure~\ref{fig:shrinkage}. The linear activation case (\( \alpha = 1 \)) is among the candidate values included in this optimization.

In the right panel, we report the corresponding predictions when we impose a linear activation function, holding all other settings fixed. It is evident from the chart that the predictions are nearly identical. This suggests that, in this case, nonlinearities do not play a substantial role in improving predictive accuracy.

It is important to emphasize that our architecture allows for flexible nonlinearities, and we select among them using out-of-sample validation. Therefore, the finding that linearity performs just as well is a \textit{genuine result}, not a consequence of model mis-specification or limitation.

In Table~\ref{tab:Table3} we report the losses of the linear model in the test and validation sample. For the convenience of the reader we report again the results for the full model, to ease comparability.

\begin{table}[htbp]
\centering
\tiny
\setlength{\tabcolsep}{1.0pt}

\begin{minipage}[t]{0.48\textwidth}
\centering
\begin{tabular}{>{\centering\arraybackslash}m{8mm} c|c|c|c|c|c|c|c}
\toprule
\multicolumn{9}{c}{\textbf{LINEAR --- VALIDATION SAMPLE}} \\
\midrule
\multicolumn{2}{c}{} & \multicolumn{7}{c}{\textbf{Quantiles ($\tau$)}} \\
\cmidrule(lr){3-9}
\multicolumn{2}{c}{} & \textbf{0.05} & \textbf{0.10} & \textbf{0.25} & \textbf{0.50} & \textbf{0.75} & \textbf{0.90} & \textbf{0.95} \\
\midrule
\multirow{11}{8mm}{\centering\rotatebox[origin=c]{90}{\textbf{Complexity}}}
& 0.0 & \cellcolor{baseline} 95.20 & \cellcolor{baseline} 81.57 & \cellcolor{baseline} 65.62 & \cellcolor{baseline} 60.82 & \cellcolor{baseline} 70.29 & \cellcolor{baseline} 89.41 & \cellcolor{baseline} 107.81 \\
& 0.1 & \cellcolor{greenlight} \begin{tabular}[c]{@{}c@{}} -1.40 \\ (5.45) \end{tabular} & \cellcolor{greenmedium} \begin{tabular}[c]{@{}c@{}} -5.26 \\ (3.33) \end{tabular} & \cellcolor{greendark} \begin{tabular}[c]{@{}c@{}} -4.40 \\ (2.27) \end{tabular} & \cellcolor{greenmedium} \begin{tabular}[c]{@{}c@{}} -3.40 \\ (2.09) \end{tabular} & \cellcolor{greendark} \begin{tabular}[c]{@{}c@{}} -7.85 \\ (3.08) \end{tabular} & \cellcolor{greendark} \begin{tabular}[c]{@{}c@{}} -17.74 \\ (3.58) \end{tabular} & \cellcolor{greendark} \begin{tabular}[c]{@{}c@{}} -24.32 \\ (6.83) \end{tabular} \\
& 0.2 & \cellcolor{greenlight} \begin{tabular}[c]{@{}c@{}} -0.29 \\ (6.90) \end{tabular} & \cellcolor{greendark} \begin{tabular}[c]{@{}c@{}} -5.82 \\ (3.50) \end{tabular} & \cellcolor{greendark} \begin{tabular}[c]{@{}c@{}} -4.55 \\ (2.46) \end{tabular} & \cellcolor{greendark} \begin{tabular}[c]{@{}c@{}} -4.24 \\ (2.48) \end{tabular} & \cellcolor{greendark} \begin{tabular}[c]{@{}c@{}} -8.98 \\ (3.35) \end{tabular} & \cellcolor{greendark} \begin{tabular}[c]{@{}c@{}} -16.62 \\ (4.42) \end{tabular} & \cellcolor{greendark} \begin{tabular}[c]{@{}c@{}} -24.37 \\ (8.67) \end{tabular} \\
& 0.3 & \cellcolor{greenlight} \begin{tabular}[c]{@{}c@{}} -0.54 \\ (7.96) \end{tabular} & \cellcolor{greenmedium} \begin{tabular}[c]{@{}c@{}} -5.17 \\ (3.82) \end{tabular} & \cellcolor{greendark} \begin{tabular}[c]{@{}c@{}} -4.34 \\ (2.83) \end{tabular} & \cellcolor{greenmedium} \begin{tabular}[c]{@{}c@{}} -3.83 \\ (2.75) \end{tabular} & \cellcolor{greendark} \begin{tabular}[c]{@{}c@{}} -9.57 \\ (3.58) \end{tabular} & \cellcolor{greendark} \begin{tabular}[c]{@{}c@{}} -17.97 \\ (4.99) \end{tabular} & \cellcolor{greendark} \begin{tabular}[c]{@{}c@{}} -23.16 \\ (10.57) \end{tabular} \\
& 0.4 & \cellcolor{redlight} \begin{tabular}[c]{@{}c@{}} 5.22 \\ (8.30) \end{tabular} & \cellcolor{greenlight} \begin{tabular}[c]{@{}c@{}} -0.19 \\ (4.70) \end{tabular} & \cellcolor{greendark} \begin{tabular}[c]{@{}c@{}} -2.58 \\ (3.55) \end{tabular} & \cellcolor{greenlight} \begin{tabular}[c]{@{}c@{}} -2.58 \\ (3.04) \end{tabular} & \cellcolor{greendark} \begin{tabular}[c]{@{}c@{}} -10.14 \\ (3.79) \end{tabular} & \cellcolor{greendark} \begin{tabular}[c]{@{}c@{}} -16.31 \\ (5.94) \end{tabular} & \cellcolor{greenlight} \begin{tabular}[c]{@{}c@{}} -16.82 \\ (13.59) \end{tabular} \\
& 0.5 & \cellcolor{reddark} \begin{tabular}[c]{@{}c@{}} 20.87 \\ (10.49) \end{tabular} & \cellcolor{redlight} \begin{tabular}[c]{@{}c@{}} 4.94 \\ (5.57) \end{tabular} & \cellcolor{greenlight} \begin{tabular}[c]{@{}c@{}} -0.42 \\ (4.06) \end{tabular} & \cellcolor{greenlight} \begin{tabular}[c]{@{}c@{}} -0.12 \\ (3.25) \end{tabular} & \cellcolor{greendark} \begin{tabular}[c]{@{}c@{}} -10.32 \\ (4.11) \end{tabular} & \cellcolor{greenmedium} \begin{tabular}[c]{@{}c@{}} -11.20 \\ (7.54) \end{tabular} & \cellcolor{greenlight} \begin{tabular}[c]{@{}c@{}} -2.68 \\ (15.44) \end{tabular} \\
& 0.6 & \cellcolor{reddark} \begin{tabular}[c]{@{}c@{}} 30.68 \\ (11.85) \end{tabular} & \cellcolor{redlight} \begin{tabular}[c]{@{}c@{}} 7.60 \\ (6.27) \end{tabular} & \cellcolor{reddark} \begin{tabular}[c]{@{}c@{}} 3.41 \\ (4.17) \end{tabular} & \cellcolor{redlight} \begin{tabular}[c]{@{}c@{}} 1.32 \\ (3.42) \end{tabular} & \cellcolor{greendark} \begin{tabular}[c]{@{}c@{}} -7.80 \\ (4.26) \end{tabular} & \cellcolor{greenlight} \begin{tabular}[c]{@{}c@{}} -5.55 \\ (8.48) \end{tabular} & \cellcolor{redlight} \begin{tabular}[c]{@{}c@{}} 13.71 \\ (16.87) \end{tabular} \\
& 0.7 & \cellcolor{reddark} \begin{tabular}[c]{@{}c@{}} 34.12 \\ (12.81) \end{tabular} & \cellcolor{redlight} \begin{tabular}[c]{@{}c@{}} 7.53 \\ (6.89) \end{tabular} & \cellcolor{reddark} \begin{tabular}[c]{@{}c@{}} 6.30 \\ (4.16) \end{tabular} & \cellcolor{reddark} \begin{tabular}[c]{@{}c@{}} 3.00 \\ (3.54) \end{tabular} & \cellcolor{greendark} \begin{tabular}[c]{@{}c@{}} -3.81 \\ (4.92) \end{tabular} & \cellcolor{greenlight} \begin{tabular}[c]{@{}c@{}} 0.04 \\ (9.18) \end{tabular} & \cellcolor{redlight} \begin{tabular}[c]{@{}c@{}} 17.17 \\ (18.45) \end{tabular} \\
& 0.8 & \cellcolor{reddark} \begin{tabular}[c]{@{}c@{}} 32.57 \\ (13.28) \end{tabular} & \cellcolor{redmedium} \begin{tabular}[c]{@{}c@{}} 10.19 \\ (7.19) \end{tabular} & \cellcolor{reddark} \begin{tabular}[c]{@{}c@{}} 9.18 \\ (3.99) \end{tabular} & \cellcolor{reddark} \begin{tabular}[c]{@{}c@{}} 4.11 \\ (3.82) \end{tabular} & \cellcolor{redlight} \begin{tabular}[c]{@{}c@{}} 0.62 \\ (5.78) \end{tabular} & \cellcolor{redlight} \begin{tabular}[c]{@{}c@{}} 4.45 \\ (10.54) \end{tabular} & \cellcolor{redlight} \begin{tabular}[c]{@{}c@{}} 23.04 \\ (20.05) \end{tabular} \\
& 0.9 & \cellcolor{reddark} \begin{tabular}[c]{@{}c@{}} 32.72 \\ (13.29) \end{tabular} & \cellcolor{redmedium} \begin{tabular}[c]{@{}c@{}} 11.92 \\ (7.33) \end{tabular} & \cellcolor{reddark} \begin{tabular}[c]{@{}c@{}} 10.48 \\ (4.07) \end{tabular} & \cellcolor{reddark} \begin{tabular}[c]{@{}c@{}} 6.07 \\ (3.88) \end{tabular} & \cellcolor{greenlight} \begin{tabular}[c]{@{}c@{}} 4.58 \\ (6.67) \end{tabular} & \cellcolor{redlight} \begin{tabular}[c]{@{}c@{}} 13.86 \\ (12.35) \end{tabular} & \cellcolor{reddark} \begin{tabular}[c]{@{}c@{}} 32.77 \\ (21.92) \end{tabular} \\
& 1.0 & \cellcolor{reddark} \begin{tabular}[c]{@{}c@{}} 37.71 \\ (12.89) \end{tabular} & \cellcolor{reddark} \begin{tabular}[c]{@{}c@{}} 20.67 \\ (7.73) \end{tabular} & \cellcolor{reddark} \begin{tabular}[c]{@{}c@{}} 12.59 \\ (4.26) \end{tabular} & \cellcolor{reddark} \begin{tabular}[c]{@{}c@{}} 6.77 \\ (4.09) \end{tabular} & \cellcolor{redlight} \begin{tabular}[c]{@{}c@{}} 7.42 \\ (7.05) \end{tabular} & \cellcolor{redmedium} \begin{tabular}[c]{@{}c@{}} 16.31 \\ (12.75) \end{tabular} & \cellcolor{reddark} \begin{tabular}[c]{@{}c@{}} 35.25 \\ (21.98) \end{tabular} \\
\bottomrule
\end{tabular}
\end{minipage}%
\hfill
\begin{minipage}[t]{0.48\textwidth}
\centering
\begin{tabular}{>{\centering\arraybackslash}m{8mm} c|c|c|c|c|c|c|c}
\toprule
\multicolumn{9}{c}{\textbf{LINEAR --- TEST SAMPLE}} \\
\midrule
\multicolumn{2}{c}{} & \multicolumn{7}{c}{\textbf{Quantiles ($\tau$)}} \\
\cmidrule(lr){3-9}
\multicolumn{2}{c}{} & \textbf{0.05} & \textbf{0.10} & \textbf{0.25} & \textbf{0.50} & \textbf{0.75} & \textbf{0.90} & \textbf{0.95} \\
\midrule
\multirow{11}{8mm}{\centering\rotatebox[origin=c]{90}{\textbf{Complexity}}}
& 0.0 & \cellcolor{baseline} 80.65 & \cellcolor{baseline} 63.69 & \cellcolor{baseline} 54.43 & \cellcolor{baseline} 54.04 & \cellcolor{baseline} 61.63 & \cellcolor{baseline} 79.56 & \cellcolor{baseline} 95.59 \\
& 0.1 & \cellcolor{greenmedium} \begin{tabular}[c]{@{}c@{}} -6.04 \\ (4.41) \end{tabular} & \cellcolor{greenlight} \begin{tabular}[c]{@{}c@{}} -0.37 \\ (2.78) \end{tabular} & \cellcolor{greenlight} \begin{tabular}[c]{@{}c@{}} -0.72 \\ (1.62) \end{tabular} & \cellcolor{greenlight} \begin{tabular}[c]{@{}c@{}} -2.32 \\ (1.85) \end{tabular} & \cellcolor{greenlight} \begin{tabular}[c]{@{}c@{}} -4.19 \\ (3.83) \end{tabular} & \cellcolor{greendark} \begin{tabular}[c]{@{}c@{}} -17.11 \\ (8.93) \end{tabular} & \cellcolor{greendark} \begin{tabular}[c]{@{}c@{}} -25.64 \\ (10.36) \end{tabular} \\
& 0.2 & \cellcolor{greenlight} \begin{tabular}[c]{@{}c@{}} -4.44 \\ (6.58) \end{tabular} & \cellcolor{redlight} \begin{tabular}[c]{@{}c@{}} 0.24 \\ (3.59) \end{tabular} & \cellcolor{greenlight} \begin{tabular}[c]{@{}c@{}} -1.04 \\ (2.49) \end{tabular} & \cellcolor{greenmedium} \begin{tabular}[c]{@{}c@{}} -3.80 \\ (2.45) \end{tabular} & \cellcolor{greenlight} \begin{tabular}[c]{@{}c@{}} -5.68 \\ (4.86) \end{tabular} & \cellcolor{greenmedium} \begin{tabular}[c]{@{}c@{}} -14.70 \\ (10.17) \end{tabular} & \cellcolor{greendark} \begin{tabular}[c]{@{}c@{}} -26.98 \\ (12.20) \end{tabular} \\
& 0.3 & \cellcolor{greenlight} \begin{tabular}[c]{@{}c@{}} -0.14 \\ (8.21) \end{tabular} & \cellcolor{redlight} \begin{tabular}[c]{@{}c@{}} 4.04 \\ (4.07) \end{tabular} & \cellcolor{greenlight} \begin{tabular}[c]{@{}c@{}} -0.19 \\ (2.95) \end{tabular} & \cellcolor{greenlight} \begin{tabular}[c]{@{}c@{}} -3.00 \\ (2.80) \end{tabular} & \cellcolor{greenlight} \begin{tabular}[c]{@{}c@{}} -6.67 \\ (5.47) \end{tabular} & \cellcolor{greenmedium} \begin{tabular}[c]{@{}c@{}} -17.13 \\ (11.00) \end{tabular} & \cellcolor{greendark} \begin{tabular}[c]{@{}c@{}} -28.23 \\ (12.29) \end{tabular} \\
& 0.4 & \cellcolor{redlight} \begin{tabular}[c]{@{}c@{}} 7.00 \\ (9.58) \end{tabular} & \cellcolor{redmedium} \begin{tabular}[c]{@{}c@{}} 6.03 \\ (4.45) \end{tabular} & \cellcolor{reddark} \begin{tabular}[c]{@{}c@{}} 2.95 \\ (3.47) \end{tabular} & \cellcolor{greenlight} \begin{tabular}[c]{@{}c@{}} -1.10 \\ (3.12) \end{tabular} & \cellcolor{greendark} \begin{tabular}[c]{@{}c@{}} -7.40 \\ (5.54) \end{tabular} & \cellcolor{greendark} \begin{tabular}[c]{@{}c@{}} -19.92 \\ (10.54) \end{tabular} & \cellcolor{greendark} \begin{tabular}[c]{@{}c@{}} -26.00 \\ (11.46) \end{tabular} \\
& 0.5 & \cellcolor{redmedium} \begin{tabular}[c]{@{}c@{}} 14.22 \\ (9.60) \end{tabular} & \cellcolor{reddark} \begin{tabular}[c]{@{}c@{}} 8.06 \\ (4.81) \end{tabular} & \cellcolor{reddark} \begin{tabular}[c]{@{}c@{}} 5.43 \\ (3.74) \end{tabular} & \cellcolor{redlight} \begin{tabular}[c]{@{}c@{}} 1.23 \\ (3.46) \end{tabular} & \cellcolor{greendark} \begin{tabular}[c]{@{}c@{}} -7.66 \\ (5.49) \end{tabular} & \cellcolor{greendark} \begin{tabular}[c]{@{}c@{}} -18.68 \\ (10.44) \end{tabular} & \cellcolor{greendark} \begin{tabular}[c]{@{}c@{}} -23.15 \\ (11.38) \end{tabular} \\
& 0.6 & \cellcolor{reddark} \begin{tabular}[c]{@{}c@{}} 21.15 \\ (10.09) \end{tabular} & \cellcolor{reddark} \begin{tabular}[c]{@{}c@{}} 11.49 \\ (5.42) \end{tabular} & \cellcolor{reddark} \begin{tabular}[c]{@{}c@{}} 8.32 \\ (3.89) \end{tabular} & \cellcolor{reddark} \begin{tabular}[c]{@{}c@{}} 2.99 \\ (3.65) \end{tabular} & \cellcolor{greendark} \begin{tabular}[c]{@{}c@{}} -7.39 \\ (6.28) \end{tabular} & \cellcolor{greenmedium} \begin{tabular}[c]{@{}c@{}} -16.08 \\ (10.65) \end{tabular} & \cellcolor{greenlight} \begin{tabular}[c]{@{}c@{}} -12.07 \\ (10.33) \end{tabular} \\
& 0.7 & \cellcolor{reddark} \begin{tabular}[c]{@{}c@{}} 30.30 \\ (11.37) \end{tabular} & \cellcolor{reddark} \begin{tabular}[c]{@{}c@{}} 19.62 \\ (6.50) \end{tabular} & \cellcolor{reddark} \begin{tabular}[c]{@{}c@{}} 9.62 \\ (3.85) \end{tabular} & \cellcolor{reddark} \begin{tabular}[c]{@{}c@{}} 4.52 \\ (3.88) \end{tabular} & \cellcolor{greendark} \begin{tabular}[c]{@{}c@{}} -5.29 \\ (6.47) \end{tabular} & \cellcolor{greenlight} \begin{tabular}[c]{@{}c@{}} -10.49 \\ (11.12) \end{tabular} & \cellcolor{greenlight} \begin{tabular}[c]{@{}c@{}} -2.68 \\ (10.36) \end{tabular} \\
& 0.8 & \cellcolor{reddark} \begin{tabular}[c]{@{}c@{}} 36.06 \\ (12.14) \end{tabular} & \cellcolor{reddark} \begin{tabular}[c]{@{}c@{}} 23.53 \\ (7.06) \end{tabular} & \cellcolor{reddark} \begin{tabular}[c]{@{}c@{}} 10.60 \\ (3.61) \end{tabular} & \cellcolor{reddark} \begin{tabular}[c]{@{}c@{}} 5.68 \\ (4.05) \end{tabular} & \cellcolor{greenlight} \begin{tabular}[c]{@{}c@{}} -2.06 \\ (6.61) \end{tabular} & \cellcolor{greenlight} \begin{tabular}[c]{@{}c@{}} -7.26 \\ (11.38) \end{tabular} & \cellcolor{redlight} \begin{tabular}[c]{@{}c@{}} 10.86 \\ (10.18) \end{tabular} \\
& 0.9 & \cellcolor{reddark} \begin{tabular}[c]{@{}c@{}} 40.39 \\ (13.14) \end{tabular} & \cellcolor{reddark} \begin{tabular}[c]{@{}c@{}} 26.67 \\ (7.66) \end{tabular} & \cellcolor{reddark} \begin{tabular}[c]{@{}c@{}} 12.22 \\ (3.65) \end{tabular} & \cellcolor{reddark} \begin{tabular}[c]{@{}c@{}} 6.74 \\ (3.80) \end{tabular} & \cellcolor{greenlight} \begin{tabular}[c]{@{}c@{}} -0.11 \\ (6.42) \end{tabular} & \cellcolor{redlight} \begin{tabular}[c]{@{}c@{}} 5.29 \\ (9.71) \end{tabular} & \cellcolor{redmedium} \begin{tabular}[c]{@{}c@{}} 17.99 \\ (12.07) \end{tabular} \\
& 1.0 & \cellcolor{reddark} \begin{tabular}[c]{@{}c@{}} 59.13 \\ (16.65) \end{tabular} & \cellcolor{reddark} \begin{tabular}[c]{@{}c@{}} 38.88 \\ (9.59) \end{tabular} & \cellcolor{reddark} \begin{tabular}[c]{@{}c@{}} 15.32 \\ (3.33) \end{tabular} & \cellcolor{reddark} \begin{tabular}[c]{@{}c@{}} 7.14 \\ (4.21) \end{tabular} & \cellcolor{redlight} \begin{tabular}[c]{@{}c@{}} 1.95 \\ (5.44) \end{tabular} & \cellcolor{redlight} \begin{tabular}[c]{@{}c@{}} 5.71 \\ (9.97) \end{tabular} & \cellcolor{reddark} \begin{tabular}[c]{@{}c@{}} 27.09 \\ (13.55) \end{tabular} \\
\bottomrule
\end{tabular}
\end{minipage}

\vspace{0.5cm}

\begin{minipage}[t]{0.48\textwidth}
\centering
\begin{tabular}{>{\centering\arraybackslash}m{8mm} c|c|c|c|c|c|c|c}
\toprule
\multicolumn{9}{c}{\textbf{DNN --- VALIDATION SAMPLE}} \\
\midrule
\multicolumn{2}{c}{} & \multicolumn{7}{c}{\textbf{Quantiles ($\tau$)}} \\
\cmidrule(lr){3-9}
\multicolumn{2}{c}{} & \textbf{0.05} & \textbf{0.10} & \textbf{0.25} & \textbf{0.50} & \textbf{0.75} & \textbf{0.90} & \textbf{0.95} \\
\midrule
\multirow{11}{8mm}{\centering\rotatebox[origin=c]{90}{\textbf{Complexity}}}
& 0.0 & \cellcolor{baseline} 95.20 & \cellcolor{baseline} 81.57 & \cellcolor{baseline} 65.62 & \cellcolor{baseline} 60.82 & \cellcolor{baseline} 70.29 & \cellcolor{baseline} 89.41 & \cellcolor{baseline} 107.81 \\
& 0.1 & \cellcolor{greenlight} \begin{tabular}[c]{@{}c@{}} -1.47 \\ (7.35) \end{tabular} & \cellcolor{greenmedium} \begin{tabular}[c]{@{}c@{}} -4.76 \\ (3.42) \end{tabular} & \cellcolor{greenmedium} \begin{tabular}[c]{@{}c@{}} -3.58 \\ (2.69) \end{tabular} & \cellcolor{greenlight} \begin{tabular}[c]{@{}c@{}} -3.43 \\ (2.68) \end{tabular} & \cellcolor{greendark} \begin{tabular}[c]{@{}c@{}} -8.33 \\ (3.33) \end{tabular} & \cellcolor{greendark} \begin{tabular}[c]{@{}c@{}} -15.10 \\ (3.92) \end{tabular} & \cellcolor{greendark} \begin{tabular}[c]{@{}c@{}} -21.44 \\ (7.56) \end{tabular} \\
& 0.2 & \cellcolor{greenlight} \begin{tabular}[c]{@{}c@{}} -0.55 \\ (8.18) \end{tabular} & \cellcolor{greenlight} \begin{tabular}[c]{@{}c@{}} -2.90 \\ (4.06) \end{tabular} & \cellcolor{greenlight} \begin{tabular}[c]{@{}c@{}} -2.77 \\ (3.60) \end{tabular} & \cellcolor{greenlight} \begin{tabular}[c]{@{}c@{}} -2.15 \\ (2.87) \end{tabular} & \cellcolor{greendark} \begin{tabular}[c]{@{}c@{}} -9.97 \\ (3.74) \end{tabular} & \cellcolor{greendark} \begin{tabular}[c]{@{}c@{}} -15.10 \\ (3.92) \end{tabular} & \cellcolor{greendark} \begin{tabular}[c]{@{}c@{}} -21.13 \\ (11.34) \end{tabular} \\
& 0.3 & \cellcolor{redlight} \begin{tabular}[c]{@{}c@{}} 0.51 \\ (8.32) \end{tabular} & \cellcolor{greenlight} \begin{tabular}[c]{@{}c@{}} -1.57 \\ (4.52) \end{tabular} & \cellcolor{greenlight} \begin{tabular}[c]{@{}c@{}} -2.59 \\ (3.98) \end{tabular} & \cellcolor{greenlight} \begin{tabular}[c]{@{}c@{}} -0.92 \\ (3.16) \end{tabular} & \cellcolor{greendark} \begin{tabular}[c]{@{}c@{}} -10.68 \\ (3.88) \end{tabular} & \cellcolor{greendark} \begin{tabular}[c]{@{}c@{}} -17.41 \\ (6.30) \end{tabular} & \cellcolor{greenlight} \begin{tabular}[c]{@{}c@{}} -13.93 \\ (13.95) \end{tabular} \\
& 0.4 & \cellcolor{redlight} \begin{tabular}[c]{@{}c@{}} 7.96 \\ (8.51) \end{tabular} & \cellcolor{redlight} \begin{tabular}[c]{@{}c@{}} 0.59 \\ (4.69) \end{tabular} & \cellcolor{redlight} \begin{tabular}[c]{@{}c@{}} 0.01 \\ (3.98) \end{tabular} & \cellcolor{redlight} \begin{tabular}[c]{@{}c@{}} 3.12 \\ (3.68) \end{tabular} & \cellcolor{greendark} \begin{tabular}[c]{@{}c@{}} -7.79 \\ (4.32) \end{tabular} & \cellcolor{greenlight} \begin{tabular}[c]{@{}c@{}} -6.58 \\ (7.53) \end{tabular} & \cellcolor{greenlight} \begin{tabular}[c]{@{}c@{}} -3.06 \\ (15.84) \end{tabular} \\
& 0.5 & \cellcolor{reddark} \begin{tabular}[c]{@{}c@{}} 24.86 \\ (11.05) \end{tabular} & \cellcolor{redlight} \begin{tabular}[c]{@{}c@{}} 4.29 \\ (5.06) \end{tabular} & \cellcolor{redmedium} \begin{tabular}[c]{@{}c@{}} 5.37 \\ (3.96) \end{tabular} & \cellcolor{redmedium} \begin{tabular}[c]{@{}c@{}} 5.76 \\ (3.94) \end{tabular} & \cellcolor{greenlight} \begin{tabular}[c]{@{}c@{}} -3.91 \\ (4.95) \end{tabular} & \cellcolor{greenlight} \begin{tabular}[c]{@{}c@{}} -3.19 \\ (7.87) \end{tabular} & \cellcolor{redlight} \begin{tabular}[c]{@{}c@{}} 14.34 \\ (17.21) \end{tabular} \\
& 0.6 & \cellcolor{reddark} \begin{tabular}[c]{@{}c@{}} 32.87 \\ (12.38) \end{tabular} & \cellcolor{reddark} \begin{tabular}[c]{@{}c@{}} 10.15 \\ (5.60) \end{tabular} & \cellcolor{redmedium} \begin{tabular}[c]{@{}c@{}} 6.32 \\ (4.16) \end{tabular} & \cellcolor{reddark} \begin{tabular}[c]{@{}c@{}} 8.39 \\ (4.07) \end{tabular} & \cellcolor{redlight} \begin{tabular}[c]{@{}c@{}} 3.41 \\ (5.76) \end{tabular} & \cellcolor{redlight} \begin{tabular}[c]{@{}c@{}} 6.52 \\ (9.56) \end{tabular} & \cellcolor{redlight} \begin{tabular}[c]{@{}c@{}} 18.22 \\ (17.43) \end{tabular} \\
& 0.7 & \cellcolor{reddark} \begin{tabular}[c]{@{}c@{}} 33.49 \\ (12.96) \end{tabular} & \cellcolor{reddark} \begin{tabular}[c]{@{}c@{}} 36.27 \\ (8.44) \end{tabular} & \cellcolor{reddark} \begin{tabular}[c]{@{}c@{}} 11.05 \\ (4.83) \end{tabular} & \cellcolor{reddark} \begin{tabular}[c]{@{}c@{}} 9.42 \\ (4.29) \end{tabular} & \cellcolor{redlight} \begin{tabular}[c]{@{}c@{}} 0.15 \\ (5.26) \end{tabular} & \cellcolor{redlight} \begin{tabular}[c]{@{}c@{}} 9.77 \\ (9.49) \end{tabular} & \cellcolor{reddark} \begin{tabular}[c]{@{}c@{}} 31.04 \\ (18.78) \end{tabular} \\
& 0.8 & \cellcolor{reddark} \begin{tabular}[c]{@{}c@{}} 34.57 \\ (14.05) \end{tabular} & \cellcolor{reddark} \begin{tabular}[c]{@{}c@{}} 37.94 \\ (8.69) \end{tabular} & \cellcolor{reddark} \begin{tabular}[c]{@{}c@{}} 18.53 \\ (5.24) \end{tabular} & \cellcolor{reddark} \begin{tabular}[c]{@{}c@{}} 10.56 \\ (4.46) \end{tabular} & \cellcolor{redlight} \begin{tabular}[c]{@{}c@{}} 6.39 \\ (6.05) \end{tabular} & \cellcolor{redmedium} \begin{tabular}[c]{@{}c@{}} 18.02 \\ (11.49) \end{tabular} & \cellcolor{reddark} \begin{tabular}[c]{@{}c@{}} 41.85 \\ (19.62) \end{tabular} \\
& 0.9 & \cellcolor{reddark} \begin{tabular}[c]{@{}c@{}} 36.19 \\ (14.21) \end{tabular} & \cellcolor{reddark} \begin{tabular}[c]{@{}c@{}} 39.35 \\ (8.07) \end{tabular} & \cellcolor{reddark} \begin{tabular}[c]{@{}c@{}} 16.61 \\ (5.05) \end{tabular} & \cellcolor{reddark} \begin{tabular}[c]{@{}c@{}} 13.58 \\ (4.64) \end{tabular} & \cellcolor{redlight} \begin{tabular}[c]{@{}c@{}} 5.52 \\ (5.82) \end{tabular} & \cellcolor{redlight} \begin{tabular}[c]{@{}c@{}} 11.28 \\ (10.48) \end{tabular} & \cellcolor{reddark} \begin{tabular}[c]{@{}c@{}} 36.76 \\ (19.38) \end{tabular} \\
& 1.0 & \cellcolor{reddark} \begin{tabular}[c]{@{}c@{}} 47.72 \\ (14.91) \end{tabular} & \cellcolor{reddark} \begin{tabular}[c]{@{}c@{}} 47.76 \\ (11.49) \end{tabular} & \cellcolor{reddark} \begin{tabular}[c]{@{}c@{}} 19.57 \\ (5.50) \end{tabular} & \cellcolor{reddark} \begin{tabular}[c]{@{}c@{}} 16.08 \\ (4.69) \end{tabular} & \cellcolor{redmedium} \begin{tabular}[c]{@{}c@{}} 8.06 \\ (5.87) \end{tabular} & \cellcolor{reddark} \begin{tabular}[c]{@{}c@{}} 20.06 \\ (11.72) \end{tabular} & \cellcolor{reddark} \begin{tabular}[c]{@{}c@{}} 39.13 \\ (20.77) \end{tabular} \\
\bottomrule
\end{tabular}
\end{minipage}%
\hfill
\begin{minipage}[t]{0.48\textwidth}
\centering
\begin{tabular}{>{\centering\arraybackslash}m{8mm} c|c|c|c|c|c|c|c}
\toprule
\multicolumn{9}{c}{\textbf{DNN --- TEST SAMPLE}} \\
\midrule
\multicolumn{2}{c}{} & \multicolumn{7}{c}{\textbf{Quantiles ($\tau$)}} \\
\cmidrule(lr){3-9}
\multicolumn{2}{c}{} & \textbf{0.05} & \textbf{0.10} & \textbf{0.25} & \textbf{0.50} & \textbf{0.75} & \textbf{0.90} & \textbf{0.95} \\
\midrule
\multirow{11}{8mm}{\centering\rotatebox[origin=c]{90}{\textbf{Complexity}}}
& 0.0 & \cellcolor{baseline} 80.65 & \cellcolor{baseline} 63.69 & \cellcolor{baseline} 54.43 & \cellcolor{baseline} 54.04 & \cellcolor{baseline} 61.63 & \cellcolor{baseline} 79.56 & \cellcolor{baseline} 95.59 \\
& 0.1 & \cellcolor{greenlight} \begin{tabular}[c]{@{}c@{}} -0.10 \\ (0.09) \end{tabular} & \cellcolor{greenlight} \begin{tabular}[c]{@{}c@{}} -0.24 \\ (2.89) \end{tabular} & \cellcolor{greenlight} \begin{tabular}[c]{@{}c@{}} -0.99 \\ (1.83) \end{tabular} & \cellcolor{greenlight} \begin{tabular}[c]{@{}c@{}} -2.52 \\ (2.08) \end{tabular} & \cellcolor{greenlight} \begin{tabular}[c]{@{}c@{}} -0.05 \\ (0.48) \end{tabular} & \cellcolor{greenmedium} \begin{tabular}[c]{@{}c@{}} -9.10 \\ (6.20) \end{tabular} & \cellcolor{greendark} \begin{tabular}[c]{@{}c@{}} -16.59 \\ (7.76) \end{tabular} \\
& 0.2 & \cellcolor{greendark} \begin{tabular}[c]{@{}c@{}} -4.50 \\ (1.25) \end{tabular} & \cellcolor{redlight} \begin{tabular}[c]{@{}c@{}} 1.65 \\ (3.69) \end{tabular} & \cellcolor{redlight} \begin{tabular}[c]{@{}c@{}} 0.26 \\ (2.83) \end{tabular} & \cellcolor{greenlight} \begin{tabular}[c]{@{}c@{}} -3.16 \\ (2.54) \end{tabular} & \cellcolor{greenlight} \begin{tabular}[c]{@{}c@{}} -6.46 \\ (5.38) \end{tabular} & \cellcolor{greenmedium} \begin{tabular}[c]{@{}c@{}} -9.10 \\ (6.20) \end{tabular} & \cellcolor{greendark} \begin{tabular}[c]{@{}c@{}} -27.46 \\ (12.30) \end{tabular} \\
& 0.3 & \cellcolor{greenlight} \begin{tabular}[c]{@{}c@{}} -1.62 \\ (6.29) \end{tabular} & \cellcolor{redlight} \begin{tabular}[c]{@{}c@{}} 3.33 \\ (4.11) \end{tabular} & \cellcolor{redlight} \begin{tabular}[c]{@{}c@{}} 2.23 \\ (3.46) \end{tabular} & \cellcolor{greenlight} \begin{tabular}[c]{@{}c@{}} -1.87 \\ (3.09) \end{tabular} & \cellcolor{greenmedium} \begin{tabular}[c]{@{}c@{}} -7.26 \\ (5.59) \end{tabular} & \cellcolor{greendark} \begin{tabular}[c]{@{}c@{}} -19.94 \\ (10.71) \end{tabular} & \cellcolor{greendark} \begin{tabular}[c]{@{}c@{}} -27.19 \\ (12.02) \end{tabular} \\
& 0.4 & \cellcolor{redlight} \begin{tabular}[c]{@{}c@{}} 7.75 \\ (9.83) \end{tabular} & \cellcolor{redlight} \begin{tabular}[c]{@{}c@{}} 5.13 \\ (4.27) \end{tabular} & \cellcolor{redlight} \begin{tabular}[c]{@{}c@{}} 4.82 \\ (4.00) \end{tabular} & \cellcolor{redlight} \begin{tabular}[c]{@{}c@{}} 0.07 \\ (3.96) \end{tabular} & \cellcolor{greenmedium} \begin{tabular}[c]{@{}c@{}} -8.17 \\ (6.13) \end{tabular} & \cellcolor{greendark} \begin{tabular}[c]{@{}c@{}} -19.50 \\ (10.72) \end{tabular} & \cellcolor{greendark} \begin{tabular}[c]{@{}c@{}} -24.05 \\ (11.52) \end{tabular} \\
& 0.5 & \cellcolor{redmedium} \begin{tabular}[c]{@{}c@{}} 14.43 \\ (9.60) \end{tabular} & \cellcolor{reddark} \begin{tabular}[c]{@{}c@{}} 10.43 \\ (4.76) \end{tabular} & \cellcolor{reddark} \begin{tabular}[c]{@{}c@{}} 8.32 \\ (4.17) \end{tabular} & \cellcolor{redlight} \begin{tabular}[c]{@{}c@{}} 1.55 \\ (3.88) \end{tabular} & \cellcolor{greenlight} \begin{tabular}[c]{@{}c@{}} -5.43 \\ (6.46) \end{tabular} & \cellcolor{greenmedium} \begin{tabular}[c]{@{}c@{}} -17.57 \\ (10.98) \end{tabular} & \cellcolor{greendark} \begin{tabular}[c]{@{}c@{}} -19.63 \\ (11.20) \end{tabular} \\
& 0.6 & \cellcolor{reddark} \begin{tabular}[c]{@{}c@{}} 21.15 \\ (10.16) \end{tabular} & \cellcolor{reddark} \begin{tabular}[c]{@{}c@{}} 14.77 \\ (5.29) \end{tabular} & \cellcolor{reddark} \begin{tabular}[c]{@{}c@{}} 9.86 \\ (4.24) \end{tabular} & \cellcolor{redlight} \begin{tabular}[c]{@{}c@{}} 3.49 \\ (3.12) \end{tabular} & \cellcolor{redlight} \begin{tabular}[c]{@{}c@{}} 0.24 \\ (5.82) \end{tabular} & \cellcolor{greenlight} \begin{tabular}[c]{@{}c@{}} -4.13 \\ (9.83) \end{tabular} & \cellcolor{greendark} \begin{tabular}[c]{@{}c@{}} -17.43 \\ (10.43) \end{tabular} \\
& 0.7 & \cellcolor{reddark} \begin{tabular}[c]{@{}c@{}} 30.01 \\ (11.35) \end{tabular} & \cellcolor{reddark} \begin{tabular}[c]{@{}c@{}} 28.15 \\ (6.55) \end{tabular} & \cellcolor{reddark} \begin{tabular}[c]{@{}c@{}} 15.62 \\ (4.02) \end{tabular} & \cellcolor{reddark} \begin{tabular}[c]{@{}c@{}} 6.48 \\ (3.16) \end{tabular} & \cellcolor{greenlight} \begin{tabular}[c]{@{}c@{}} -2.67 \\ (6.18) \end{tabular} & \cellcolor{greenlight} \begin{tabular}[c]{@{}c@{}} -5.90 \\ (9.33) \end{tabular} & \cellcolor{redlight} \begin{tabular}[c]{@{}c@{}} 5.07 \\ (11.10) \end{tabular} \\
& 0.8 & \cellcolor{reddark} \begin{tabular}[c]{@{}c@{}} 37.47 \\ (12.15) \end{tabular} & \cellcolor{reddark} \begin{tabular}[c]{@{}c@{}} 25.52 \\ (6.98) \end{tabular} & \cellcolor{reddark} \begin{tabular}[c]{@{}c@{}} 21.16 \\ (3.91) \end{tabular} & \cellcolor{reddark} \begin{tabular}[c]{@{}c@{}} 9.70 \\ (3.21) \end{tabular} & \cellcolor{redlight} \begin{tabular}[c]{@{}c@{}} 1.36 \\ (5.64) \end{tabular} & \cellcolor{redlight} \begin{tabular}[c]{@{}c@{}} 1.22 \\ (8.29) \end{tabular} & \cellcolor{redlight} \begin{tabular}[c]{@{}c@{}} 15.74 \\ (13.11) \end{tabular} \\
& 0.9 & \cellcolor{reddark} \begin{tabular}[c]{@{}c@{}} 39.29 \\ (12.66) \end{tabular} & \cellcolor{reddark} \begin{tabular}[c]{@{}c@{}} 23.92 \\ (5.77) \end{tabular} & \cellcolor{reddark} \begin{tabular}[c]{@{}c@{}} 21.23 \\ (4.08) \end{tabular} & \cellcolor{reddark} \begin{tabular}[c]{@{}c@{}} 17.74 \\ (3.48) \end{tabular} & \cellcolor{redlight} \begin{tabular}[c]{@{}c@{}} 0.79 \\ (5.52) \end{tabular} & \cellcolor{greenlight} \begin{tabular}[c]{@{}c@{}} -3.16 \\ (9.01) \end{tabular} & \cellcolor{redlight} \begin{tabular}[c]{@{}c@{}} 12.04 \\ (11.92) \end{tabular} \\
& 1.0 & \cellcolor{reddark} \begin{tabular}[c]{@{}c@{}} 66.45 \\ (16.78) \end{tabular} & \cellcolor{reddark} \begin{tabular}[c]{@{}c@{}} 24.08 \\ (6.66) \end{tabular} & \cellcolor{reddark} \begin{tabular}[c]{@{}c@{}} 23.27 \\ (4.79) \end{tabular} & \cellcolor{reddark} \begin{tabular}[c]{@{}c@{}} 27.20 \\ (4.94) \end{tabular} & \cellcolor{redlight} \begin{tabular}[c]{@{}c@{}} 2.53 \\ (5.78) \end{tabular} & \cellcolor{redlight} \begin{tabular}[c]{@{}c@{}} 4.98 \\ (8.31) \end{tabular} & \cellcolor{redmedium} \begin{tabular}[c]{@{}c@{}} 28.56 \\ (19.21) \end{tabular} \\
\bottomrule
\end{tabular}
\end{minipage}

\vspace{0.3cm}
\caption{Accuracy of the forecast of the change in the unemployment rate 1 month ahead. The first row shows absolute loss of naive model based on recursively estimated unconditional quantiles, normalized by $\tau(1-\tau)$. Subsequent rows show differences of model loss relative to naive baseline, normalized by $\tau(1-\tau)$. Negative values (green) indicate model outperforms naive; positive values (red) indicate underperformance. HAC standard errors in parentheses (robust to autocorrelation and heteroskedasticity), also normalized by $\tau(1-\tau)$. Color darkness based on $|\text{differential}/\text{standard error}|$ ratio using thresholds 1.28 and 1.65 (90th and 95th quantiles of the normal distribution): light shading below 1.28, medium between 1.28 and 1.65, darker if larger than 1.65.}
\label{tab:Table3}
\end{table}

We conclude that the model is able to detect macroeconomic vulnerabilities as they emerge. However, in this setting, the empirical evidence indicates that nonlinear transformations are not essential for forecasting accuracy.

\subsubsection{Robustness to Outliers}

Forecasts based on the pinball loss function are generally more robust than those based on the quadratic loss. This difference stems from how these loss functions treat extreme observations. The quadratic loss assigns disproportionate weight to large errors, making it highly sensitive to outliers. In contrast, the pinball loss increases linearly with the error, thus downweighting extreme values and enhancing robustness.

\begin{figure}[H]
  \centering
  \begin{subfigure}[b]{0.40\textwidth}
    \centering
    \includegraphics[width=\linewidth]{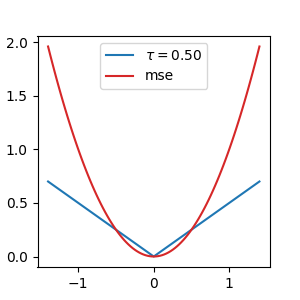}
    \caption{Pinball loss ($\tau=0.5$) vs MSE}
    \label{fig:pinball-mse}
  \end{subfigure}\hfill
  \begin{subfigure}[b]{0.55\textwidth}
    \centering
    \includegraphics[width=\linewidth]{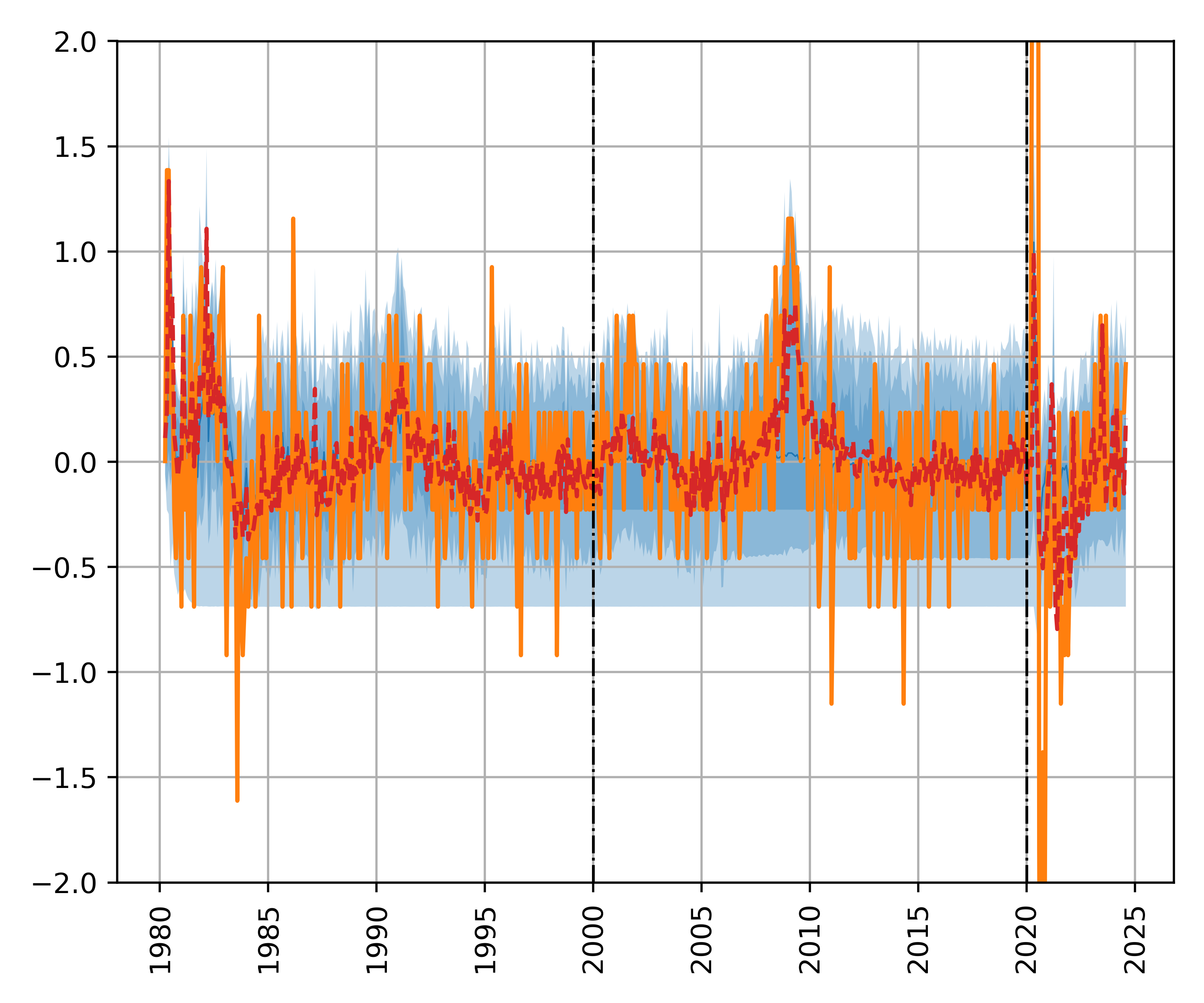}
    \caption{Forecast and OLS comparison}
    \label{fig:forecast-ols}
  \end{subfigure}
    \caption{(a) Pinball loss with \( \tau=0.5 \) (blue) and MSE (red); 
    (b) Forecast of the change in the unemployment rate one month ahead: Actual (solid, orange) and quantiles of the predicted distribution (shaded, blue), and OLS (dashed, red).}
  \label{fig:prediction_mse}
\end{figure}

This distinction is well-known in the context of estimating location parameters: the median (which minimizes the pinball loss for \( \tau = 0.5 \)) is more robust than the mean (which minimizes the quadratic loss). The left panel of Figure~\ref{fig:prediction_mse} illustrates this point. For a given observed outcome, large deviations from the predicted value have a much larger impact on the quadratic loss than on the pinball loss.

The right panel of Figure~\ref{fig:prediction_mse} provides supporting evidence from our empirical application. After the COVID-19 outbreak, predictions based on the pinball loss remain stable and plausible, while predictions based on the quadratic loss become erratic and unstable. This underscores the empirical value of quantile-based forecasting, particularly in turbulent periods.

\subsubsection{Additional Evidence: Other Variables and Horizons}

\begin{figure}[H]
  \centering
  \begin{subfigure}[b]{0.45\textwidth}
    \centering
    \includegraphics[width=\linewidth]{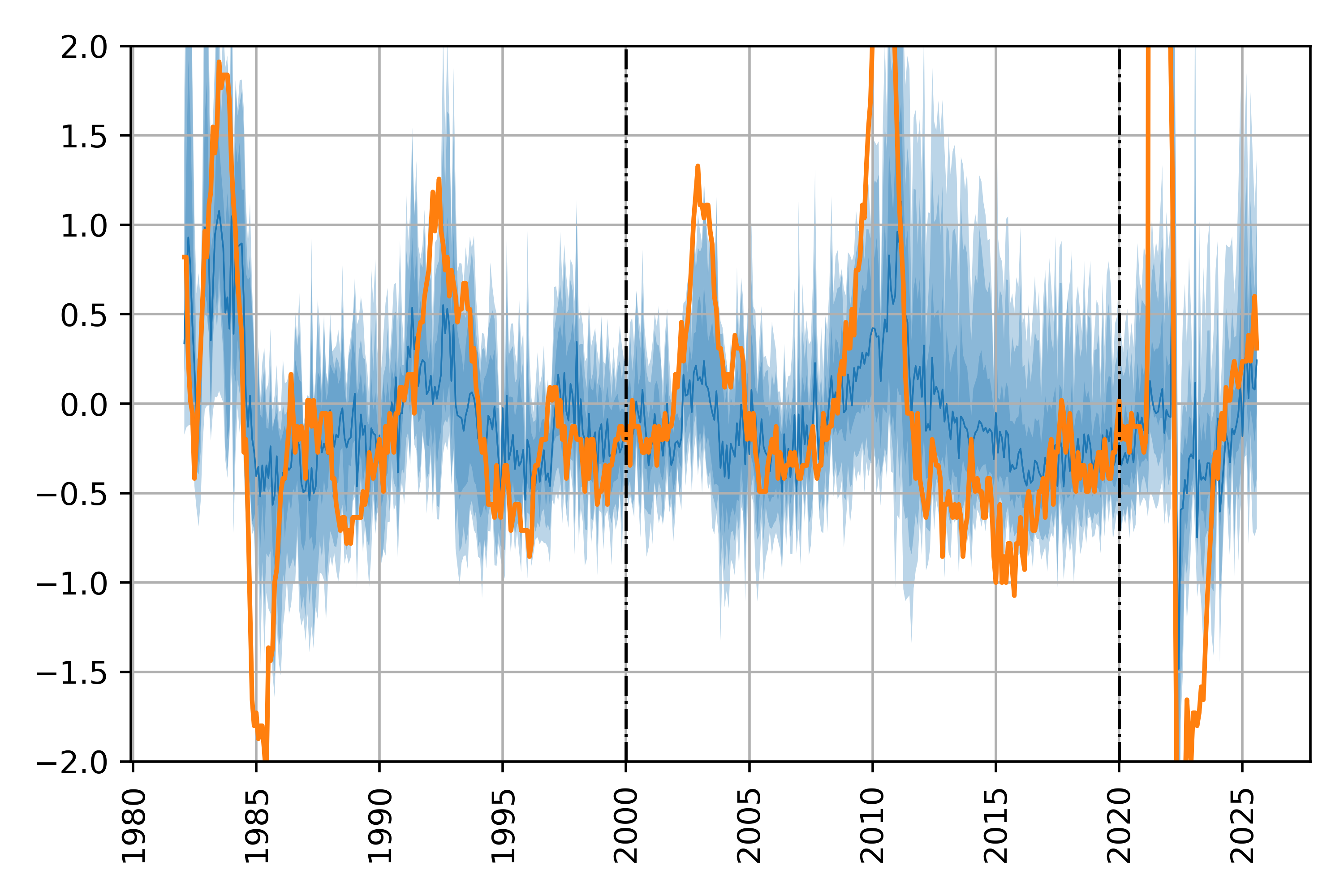}
    \caption{Leaky ReLU}
  \end{subfigure}\hfill
  \begin{subfigure}[b]{0.45\textwidth}
    \centering
    \includegraphics[width=\linewidth]{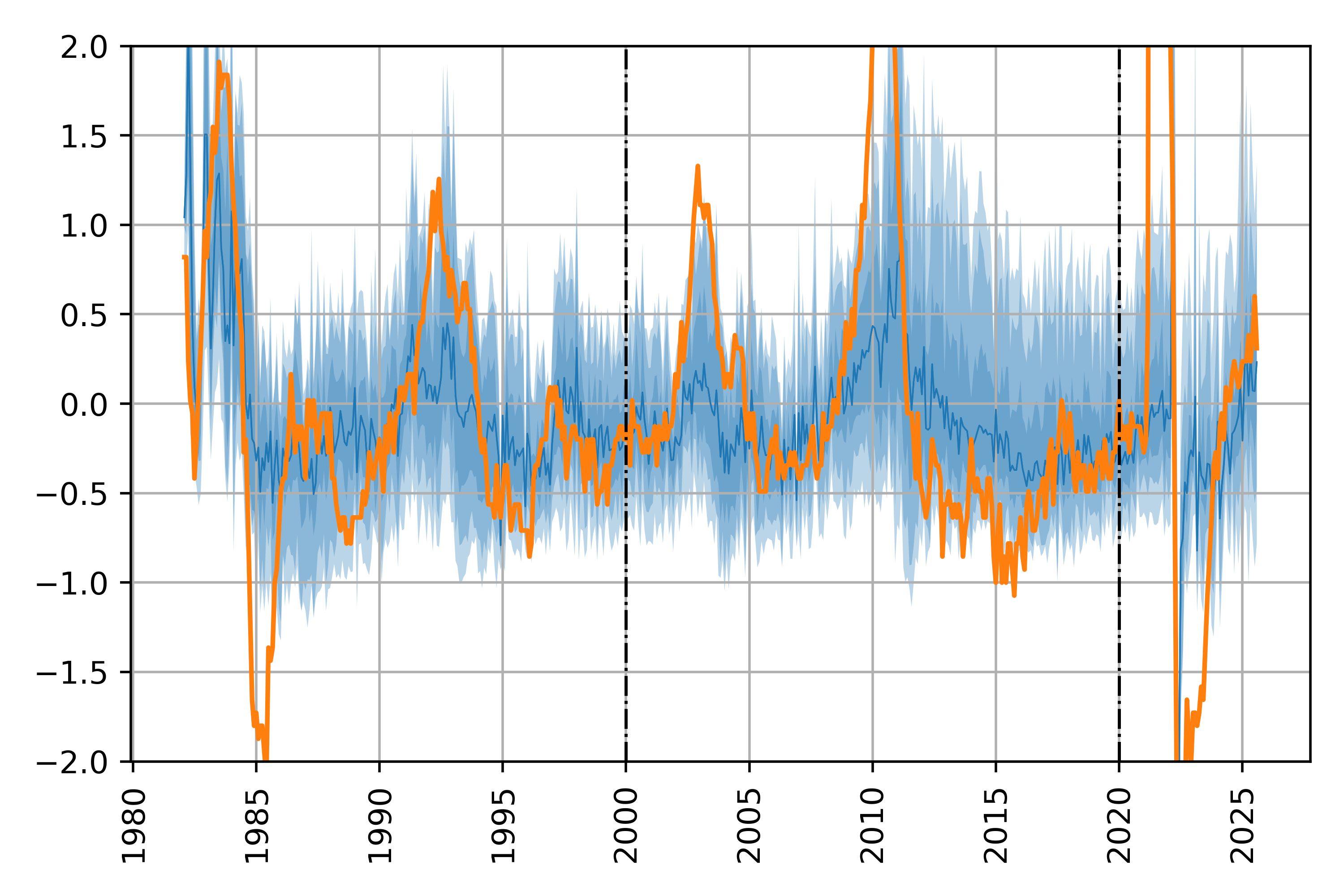}
    \caption{Linear}
  \end{subfigure}
 \caption{Forecast of the year-over-year change in the unemployment one year ahead. Actual (solid, orange) and quantiles of the predicted distribution (shaded, blue) with:  (a) Leaky ReLU activation (b) Linear activation.}
   \label{fig:unrate_12m}
\end{figure}

The baseline exercise has focused on forecasting unemployment one month ahead. Interestingly, the main conclusions are valid for longer horizons and other target variables. We explored many combinations and found that these insights consistently hold. Comprehensive results are omitted for brevity but are available upon request. As illustrative examples, we report forecasts for the year-over-year change in unemployment one year ahead (Figure~\ref{fig:unrate_12m}).

The model captures upside macroeconomic risk at longer horizons, detecting predictable movements in the upper quantiles of unemployment. This confirms that the general approach is effective even when forecasting risk further into the future.

\begin{figure}[H]
  \centering
  \begin{subfigure}[b]{0.45\textwidth}
    \centering
    \includegraphics[width=\linewidth]{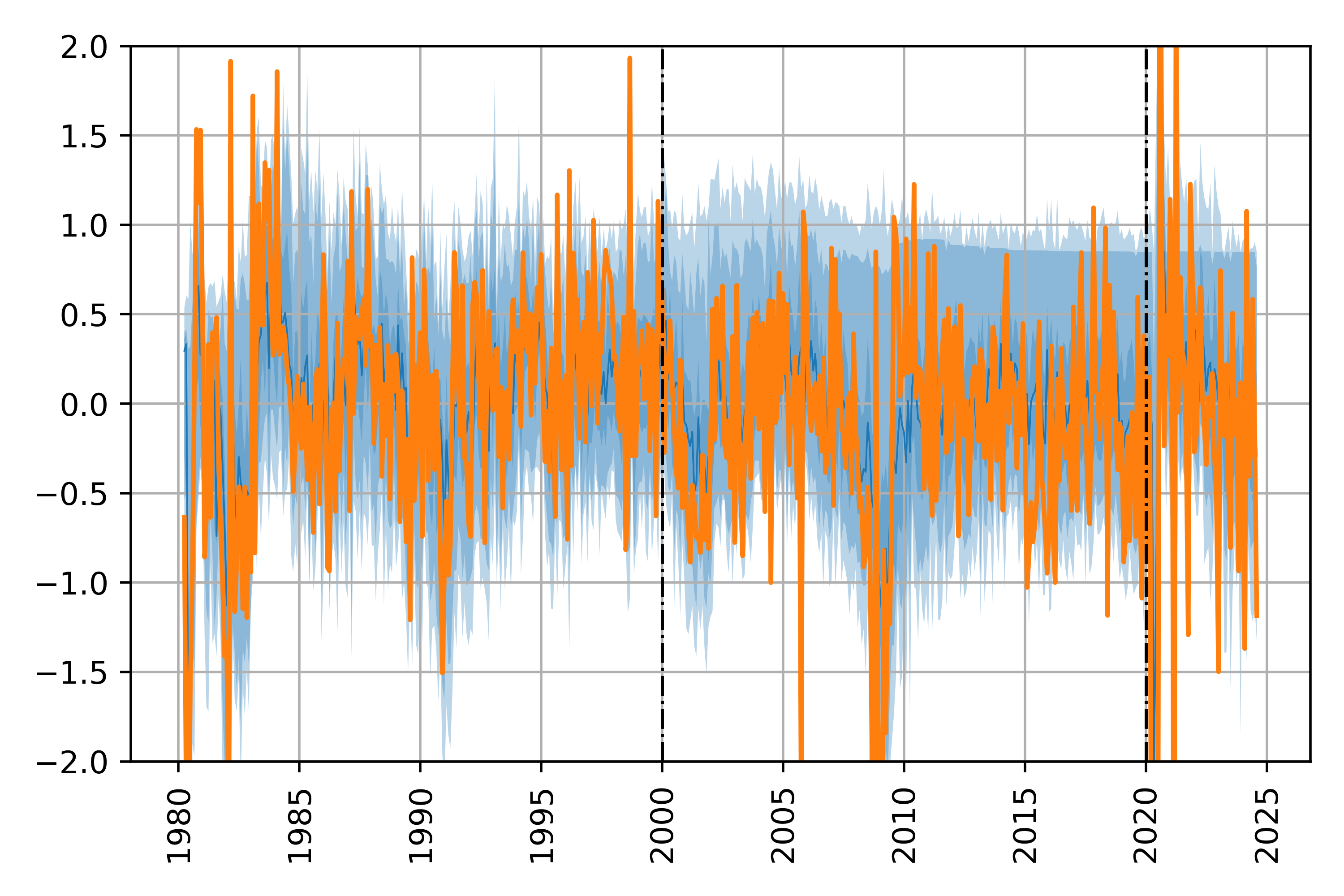}
    \caption{Leaky ReLU}
  \end{subfigure}\hfill
  \begin{subfigure}[b]{0.45\textwidth}
    \centering
    \includegraphics[width=\linewidth]{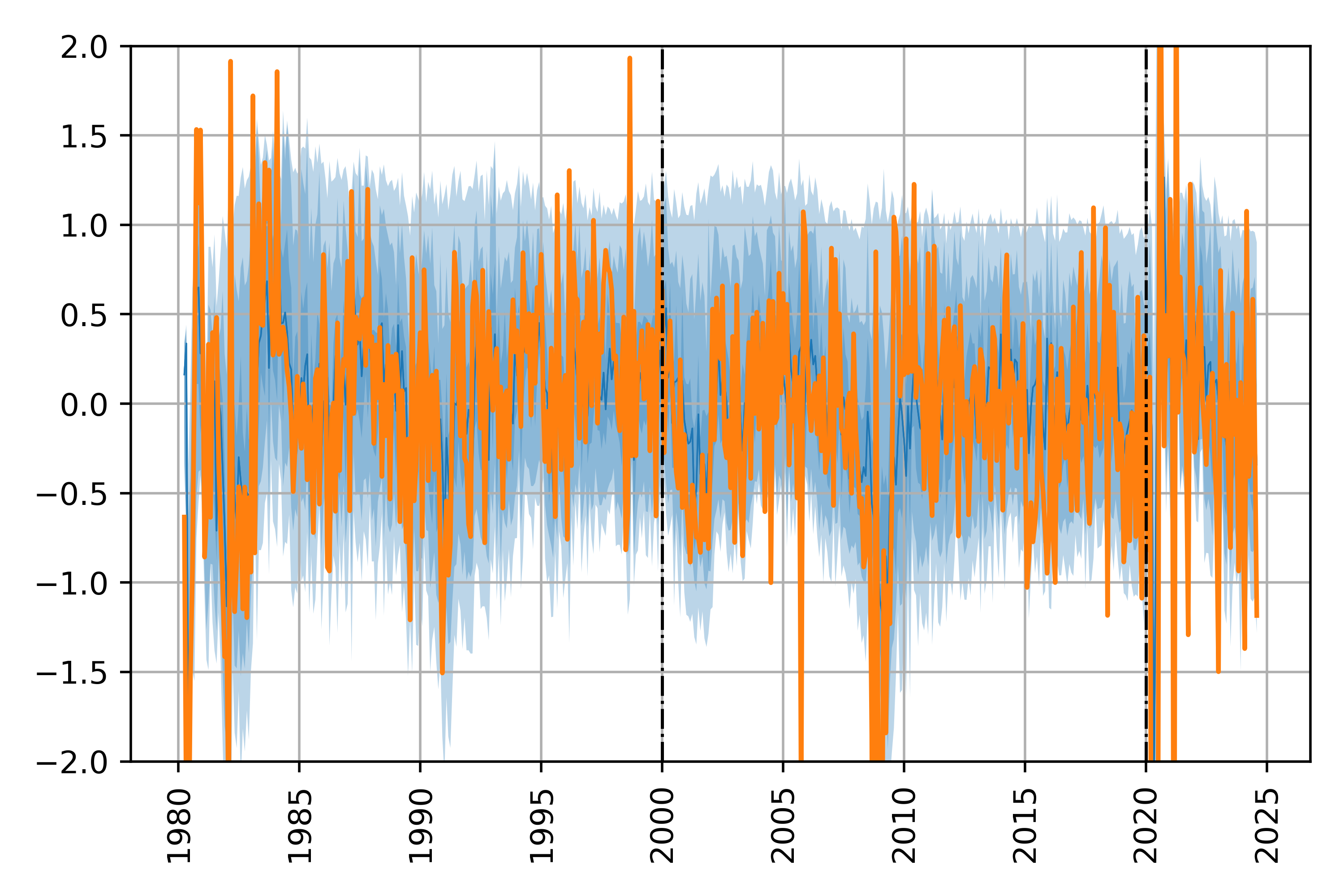}
    \caption{Linear}
  \end{subfigure}
    \caption{Forecast of the year-over-year change in the unemployment one year ahead with: (a) Leaky ReLU (b) Linear activation}  \label{fig:indpro_1m}
\end{figure}

We also report the forecast of the monthly growth rate of industrial production one month ahead (Figure~\ref{fig:indpro_1m}).

The model detects downside macroeconomic risk by identifying predictable declines in the lower quantiles of industrial production growth. 

In both cases, the forecasts generated by the full DNN and those with linear activation functions are nearly identical. The generality of this empirical result echoes \cite{CarrieroEtal2024}, who use Time-Series Large Language Models (TS-LLMs) for macroeconomic forecasting with Big Data. These models adapt transformer-based large language models, originally developed for text, to time-series forecasting. Using a similar dataset, they find that TS-LLMs do not outperform linear factor models, which are closely related to principal components  \citep[see][]{DozEtal2012}, or large Bayesian VARs, which are closely related to linear ridge regression \citep[see][]{BanburaEtal2010, BanburaEtal2015}.

\section{Conclusions}
\label{sec:conclusion}

In summary, the results above show that the model is able to capture business cycle vulnerabilities as they emerge by exploiting the information contained in Big Data through a highly flexible and general predictive model. There are no major variations in upside macroeconomic risk, while substantial variation is observed on the downside. Methodologically, our conclusion is that machine learning models, when properly regularized and validated, can robustly uncover predictive structures in high-dimensional macroeconomic data, even when their functional complexity is not ultimately required. We have shown that using out-of-sample performance to guide architecture and hyperparameter selection is essential. Otherwise, the signal in the data would be lost in the noise due to overfitting. Thanks to this validation-based selection process, the complexity of the DNN does not lead to overfitting. In this context, shrinkage is crucial. Allowing for nonlinearities does not provide substantial gains in predictive accuracy, in this case.

Throughout the paper, we focused on partial models, in which the objective was to predict one variable at a time. The natural extension is to full models, where all variables are predicted jointly. In the linear setting, the literature has built up developing large Bayesian vector autoregressions (VARs) that leverage regularization techniques with a direct connection to informative prior specifications, and shown that they deliver accurate point and density forecasts while offering both computational tractability and theoretical grounding for inference in high dimension \citep{BanburaEtal2010,Koop2013JAE,GiannoneEtal2015}. Related contributions incorporated stochastic volatility to further enhance density forecasting performance \citep[e.g.,][]{Clark2011JBES,Chan2020JBES}. In the more general setting of this paper, however, the extension was far from trivial: the multivariate counterpart of quantile predictions posed conceptual challenges, since there is no natural ordering in multiple dimensions, and the simulation of nonlinear models to recover longer horizon forecasts required particular care. Recent work by \citet{AdrianEtal2019} explored nonparametric multivariate distribution regression models, but their approach was limited to a handful of variables and did not easily extend to high-dimensional settings. We therefore leave the development of such high-dimensional nonlinear multivariate models for future research. At the same time, the results of this paper provided a useful guide in that direction: in particular, the finding that linear models performed well at the quantile level offered a natural starting point for developing tractable high-dimensional multivariate frameworks. Such extensions would be crucial for advancing risk assessment and policy analysis, as understanding the joint distribution of many variables is essential for monitoring vulnerabilities and the buildup of risks in the economy.

While this paper focuses on forecasting macroeconomic risks in the United States, the framework developed here—evaluating the role of regularization through shrinkage, model architecture choices such as depth and activation functions, and the use of out-of-sample validation—can be applied more broadly. Future work could extend this analysis to other economic domains, including finance, microeconomic applications, and international macroeconomics, as well as to different data structures such as cross-sectional and panel data, in the spirit of \cite{GiannoneEtal2021}.

\newpage 
\small
\bibliography{DNN}
\bibliographystyle{chicago} 
\normalsize

\end{document}